\documentclass[final]{siamltex}

\usepackage{amsfonts,amsmath}
\usepackage[dvips]{epsfig}
\usepackage{graphics,graphicx}
\usepackage{showkeys} 
\usepackage{mcode}

\makeatletter\@addtoreset {equation}{section}\makeatother

\title{Families of Surface Gap Solitons and their Stability via the Numerical Evans Function Method}

\author{Elizabeth Blank\thanks{DFG Research Training Group 1294 Analysis, Simulation and Design of Nanotechnological Processes, Karlsruhe Institute of Technology, Karlsruhe, Germany.} \and Tom\'{a}\v{s} Dohnal\thanks{DFG Research Training Group 1294 Analysis, Simulation and Design of Nanotechnological Processes, Karlsruhe Institute of Technology, Karlsruhe, Germany ({\tt dohnal@kit.edu}).}}

\begin{document}

\maketitle

\def\bexe{\begin{exercise}}\def\eexe{\eex\end{exercise}}
\def\bsol{\begin{solution}}\def\esol{\eex\end{solution}}
\def\bexa{\begin{example}}\def\eexa{\eex\end{example}}
\def\brem{\begin{remark}}\def\erem{\end{remark}}
\def\bthm{\begin{theorem}}\def\ethm{\end{theorem}}
\def\blem{\begin{lemma}}\def\elem{\end{lemma}}
\def\bcor{\begin{corollary}}\def\ecor{\end{corollary}}
\def\beq{\begin{equation}}\def\eeq{\end{equation}}
\def\i{\item}

\newcommand{\R}{{\mathbb R}}
\newcommand{\C}{{\mathbb C}}\newcommand{\N}{{\mathbb N}}\renewcommand{\P}{{\mathbb P}}
\newcommand{\Q}{{\mathbb Q}}\newcommand{\T}{{\mathbb T}}\newcommand{\Z}{{\mathbb Z}}

\def\L{{\mathbb L}}
\def\CC{{\cal C}}\def\CK{{\cal K}}\def\CT{{\cal T}}\def\CU{{\cal U}}
\def\CV{{\cal V}}\def\CA{{\cal A}}\def\CE{{\cal E}}\def\CD{{\cal D}}  
\def\CF{{\cal F}}\def\CG{{\cal G}}\def\CH{{\cal H}}\def\CL{{\cal L}}
\def\CO{{\cal O}}\def\CP{{\cal P}}\def\CR{{\cal R}}\def\CS{{\cal S}}
\def\CM{{\cal M}}\def\CN{{\cal N}}\def\CZ{{\cal Z}}\def\CB{{\cal B}}
\def\CX{{\cal X}}\def\CY{{\cal Y}}

\def\Bsm{{\scriptscriptstyle B}}

\def\ga{\gamma}\def\om{\mu}\def\th{\theta}\def\uh{\hat{u}}\def\vh{\hat{v}}
\def\noi{\noindent}\def\ds{\displaystyle}
\def\vt{\vartheta}\def\Id{\mbox{Id}}\def\pa{{\partial}}\def\lam{\lambda}
\newcommand{\bi}{\begin{itemize}}\newcommand{\ei}{\end{itemize}}
\newcommand{\ben}{\begin{enumerate}}\newcommand{\een}{\end{enumerate}}
\newcommand{\bce}{\begin{center}}\newcommand{\ece}{\end{center}}
\newcommand{\reff}[1]{(\ref{#1})}\newcommand{\ul}[1]{{\underline {#1}}}
\newcommand{\mbf}[1]{{\mathbf{#1}}}\newcommand{\ov}[1]{{\overline {#1}}}
\newcommand{\ol}[1]{{\overline {#1}}}\newcommand{\nonu}{\nonumber}
\newcommand{\spr}[1]{\left\langle #1 \right\rangle}
\newcommand{\hs}[1]{{\hspace{#1}}}\newcommand{\vs}[1]{{\vspace{#1}}}
\newcommand{\bhs}[1]{\mbox{\hspace{#1}}}
\def\eps{\varepsilon}\def\epsi{\epsilon}\def\aqui{\Leftrightarrow}
\def\ra{\rightarrow}\def\Ra{\Rightarrow}\def\la{\leftarrow}
\def\lra{\leftrightarrow}\def\lora{\longrightarrow}\def\loma{\longmapsto}
\newcommand{\barr}{\begin{array}}\newcommand{\earr}{\end{array}}
\newcommand{\bpm}{\begin{pmatrix}}\newcommand{\epm}{\end{pmatrix}}
\newcommand{\ba}{\begin{array}}\newcommand{\ea}{\end{array}}
\def\dd{\, {\rm d}}\def\ri{{\rm i}}\def\sech{{\rm sech}}
\def\dx{\, {\rm d}x}
\def\cc{{\rm c.c.}}\def\res{{\rm Res}}\def\er{{\rm e}}
\def\re{{\rm Re}}\def\im{{\rm Im}}\def\ccf{{\rm c.c.f}}
\def\spec{{\rm spec}}\def\om{\omega}\def\Om{\Omega}
\def\hot{{\rm h.o.t}}\def\ddt{\frac{\rm d}{{\rm d}t}}
\def\ddx{\frac{\rm d}{{\rm d}x}}\def\bphi{\varphi}\def\del{\delta}
\def\supp{{\rm supp}}\def\id{{\rm Id}}

\def\nab{\nabla}\def\eex{\hfill\mbox{$\rfloor$}}
\def\Del{\Delta}\def\na{\nabla}
\def\phiti{\tilde{\phi}}\def\psiti{\tilde{\psi}}
\def\kti{\tilde{k}}\def\sig{\sigma}
\def\nag{\nabla{\gamma}}\def\lap{\Delta}\def\al{\alpha}
\def\fti{\tilde{f}}\def\gti{\tilde{g}}\def\wti{\tilde{w}}
\def\he{\hat{e}}\def\hemu{\hat{e}_{(\mu)}}\def\henu{\hat{e}_{(\nu)}}
\def\hemp{\hat{e}_{(\mu')}}\def\henp{\hat{e}_{(\nu')}}
\def\mup{{\mu'}}\def\nup{{\nu'}}
\def\wh{\widehat}

\def\Ga{\Gamma}\def\Rti{\tilde{R}}\def\Lam{\Lambda}
\def\kap{\kappa}\def\Sig{\Sigma}
\def\rot{\nabla\times}\def\nati{\nabla\times}\def\nad{\nabla\cdot}

\def\xmin{{x_{\mbox{\small min}}}}\def\xmax{{x_{\mbox{\small max}}}} 
\def\un{u^{(0)}}\def\ue{u^{(1)}} \def\us{{\tilde{u}}}
\def\bd{\begin{displaymath}} \def\ed{\end{displaymath}}
\def\ba{\begin{array}} \def\ea{\end{array}}
\def\ua{u^{(1)}} \def\ub{u^{(2)}}\def\uta{u_t^{(1)}} \def\utb{u_t^{(2)}}
\def\eps{\varepsilon} \def\dt{\Delta t \;}
\def\lp{\Delta} \def\cd{(t,\cdot)}\def\cn{(0,\cdot)} \def\di{\displaystyle}
\def\xppaut{{\tt xppaut}}\def\xpp{{\tt xppaut}}



\begin{abstract}
The nonlinear Schr\"{o}dinger/Gross-Pitaevskii equation with a linear periodic potential and a nonlinearity coefficient $\Gamma$ with a discontinuity supports stationary localized solitary waves with frequencies inside spectral gaps, so called surface gap solitons (SGSs). We compute families of 1D SGSs using the arclength continuation method for a range of values of the jump in $\Gamma$. 
Using asymptotics, we show that when the frequency parameter converges to the bifurcation gap edge, the size of the allowed jump in $\Gamma$ converges to $0$ for SGSs centered at any $x_c\in \R$.

Linear stability of SGSs is next determined via the numerical Evans function method, in which the stable and unstable manifolds corresponding to the $0$ solution of the linearized spectral ODE problem are evolved up to a common location where the determinant of their bases, i.e., the Evans function, is evaluated. Zeros of the Evans function coincide with eigenvalues of the linearized operator. Far from the SGS location the manifolds are spanned by exponentially decaying/increasing Bloch functions. As we show, evolution of the manifolds suffers from stiffness. A numerically stable formulation is possible in the exterior algebra formulation and with the  use of Grassmanian preserving ODE integrators. Eigenvalues with a positive real part larger than a small constant
are then detected via the use of the complex argument principle and a contour parallel to the imaginary axis. The location of real eigenvalues is found via a straightforward evaluation of the Evans function along the real axis and several complex eigenvalues are located using M\"{u}ller's method. The numerical Evans function method is described in detail in order to facilitate its use as a practical tool for locating eigenvalues. Our results show the existence of both unstable and stable SGSs (possibly with a weak instability), where stability is obtained even for some SGSs centered in the domain half with the less focusing nonlinearity. Direct simulations of the PDE for selected SGS examples confirm the results of Evans function computations.
\end{abstract}



\pagestyle{myheadings}
\thispagestyle{plain}
\markboth{E. BLANK AND T. DOHNAL}
{\textsc{Surface Gap Solitons; Stability via the Numerical Evans Function}}

\section{Introduction}\label{S:intro}
Waves in periodic media with a nonlinear response are of fundamental importance in several areas of physics including propagation of light in nonlinear photonic crystals and evolution of Bose Einstein condensates (BECs) loaded on optical lattices. A classical and highly universal one dimensional model in these settings is the periodic nonlinear Schr\"{o}dinger (PNLS) / Gross-Pitaevskii equation 
\begin{align}\label{E:PNLS}
\ri \partial_t \psi(x,t) + \partial_x^2 \psi(x,t) - V(x) \psi(x,t) + \Gamma(x) |\psi(x,t)|^2 \psi(x,t)=0 
\end{align}
on $x\in \R, t\geq 0$ with $V(x+d)=V(x) \ \forall \; x \in\R$. Equation \eqref{E:PNLS} is a model for BECs confined via a magnetic or optical trap (in the directions transversal to $x$) to a quasi-1D geometry and loaded on an optical lattice
\cite{DGPS99,EC03} and also for optical beams propagating in Kerr nonlinear media in a direction orthogonal to the $x-$axis along which the linear part of the refractive index of the medium is periodic \cite{SK02}. In the latter case $t$ is a spatial variable along the propagation direction.  We investigate stationary solitary waves for nonlinear interfaces
\beq\label{E:Gamma}
\Gamma(x) = \Gamma_{+} \; \chi_{[0,\infty)}(x) + \Gamma_{-} \; \chi_{(-\infty,0)}(x), \ \text{where} \ \Gamma_\pm\in\R.
\eeq
In optics this coefficient describes two media with different values of the cubic susceptibility affixed at $x=0$ while in BECs it corresponds to two condensates with different $s$-wave scattering lengths.
If $\Gamma(x)>0$, the BEC interaction is `attractive' at $x$; in optics the medium is called `focusing' at $x$. In the opposite case the respective terms are `repulsive' and `defocusing'.

Localized solitary waves are of particular physical as well as mathematical interest due to their properties of finite energy and constant shape throughout evolution. In optics they are considered as viable candidates for bit carriers. Solitary wave solutions $\psi(x,t) = e^{-\ri \mu t} \phi(x), \mu\in \R$ of \eqref{E:PNLS} with a periodic $V$ and with exponentially localized $\phi$ are called \textit{surface gap solitons} (SGSs) when $\Gamma_+\neq \Gamma_-$ in \eqref{E:Gamma} and \textit{gap solitons} (GSs) when $\Gamma$ is periodic. The spatial profile $\phi$ satisfies the stationary periodic nonlinear Schr\"{o}dinger equation
\begin{align}\label{E:SPNLS_cplx}
\phi''(x) + \mu\phi(x) -  V(x) \phi(x) + \Gamma(x) |\phi(x)|^2\phi(x)=0, \quad x\in \R.
\end{align}
The word `gap' in the names of GS and SGS comes from the fact that the frequency parameter $\mu$ must lie in a gap of the spectrum $\sigma(L)$ of the operator
\[L:=-\pa_x^2+V(x), \ x\in \R.\]
(S)GSs arise due to a balance between the linear periodicity induced dispersion and the nonlinear focusing or defocusing. GSs of \eqref{E:PNLS} were studied theoretically, for instance, in \cite{SK02,PSK04}. In \cite{Stuart95,Pankov05} existence of gap soliton ground states in all gaps of $\sigma(L)$ for the $d$-dimensional PNLS was proved rigorously using variational methods. GSs have been also observed experimentally in optics \cite{ChLS03,FSECh03} as well as in BECs \cite{Eiermann_etal_04}.

One dimensional SGSs have been previously studied mainly at linear interfaces, i.e., for $\Gamma \equiv \text{const.}$ and $V(x) = V_1(x) \; \chi_{[0,\infty)}(x) + V_2(x) \; \chi_{(-\infty,0)}(x), \ V_{1,2}(x+d_{1,2})=V_{1,2}(x)$, see the numerical studies e.g. in \cite{KVT06,MHChSMS06}. Analogous linear interfaces have been considered in 2D \cite{KEVT06,KT06}. SGSs have been observed also experimentally in photonic crystals \cite{Rosberg_etal_06,Szameit_etal_08}.

At the nonlinear interface \eqref{E:Gamma} SGSs were studied via numerical and asymptotic methods in \cite{DP08}. Using the implicit function theorem, their existence was shown to hold via a parameter continuation from GSs with $\Gamma \equiv$const. for sufficiently small $|\Gamma_+ - \Gamma_-|$ under a non-degeneracy condition on the GS. Families of SGSs were then numerically continued in $\Gamma_-$ for $\Gamma_+$ fixed up to a critical point where the corresponding Jacobian becomes singular. Here we first  use the numerical arclength continuation to extend these families past the critical points, which are all apparently simple folds of the SGS families. The families are continued up to a chosen threshold value of $\Gamma_-$ or total energy. Next, we investigate linear stability of these SGSs via the numerical Evans function method and identify stable and unstable segments of the SGS families. We show that folds  are often associated with a stability change. 

The phenomenon of localization in a medium with a purely nonlinear interface relies crucially on the presence of a linear structure. Without a linear structure, i.e., for $V\equiv$ const., no localized solutions of \eqref{E:SPNLS_cplx} exist as shown in the following lemma. This result also appears (for real $\phi$) in Sec. 3.3 of \cite{DP08}.
\begin{lemma}\label{L:nonexistence_homog}
If $V\equiv \text{const.}$ and $\Gamma_+\neq \Gamma_-$, then the stationary periodic nonlinear Schr\"{o}dinger equation \eqref{E:SPNLS_cplx} has no nontrivial solutions in $H^1(\R)$. 
\end{lemma}
\begin{proof}
Due to the presence of the term $\mu \phi$ in \eqref{E:SPNLS_cplx} we can assume without any loss of generality that $V\equiv 0$. As a first order system in real variables $\phi_\text{Re}:=\text{Re}(\phi)$ and $\phi_\text{Im}:=\text{Im}(\phi)$ equation \eqref{E:SPNLS_cplx} reads
\[
\frac{d}{dx}\left(\begin{smallmatrix}\phi_\text{Re}\\ \phi_\text{Im}\\ \phi'_\text{Re}\\ \phi'_\text{Im}\end{smallmatrix}\right) = \left(\begin{smallmatrix}\phi'_\text{Re}\\ \phi'_\text{Im}\\ -\mu \phi_\text{Re} -\Gamma(x)(\phi^2_\text{Re}+\phi^2_\text{Im})\phi_\text{Re}\\ -\mu \phi_\text{Im} -\Gamma(x)(\phi^2_\text{Re}+\phi^2_\text{Im})\phi_\text{Im}\end{smallmatrix}\right),
\]
which is a Hamiltonian  system with $H(\phi_\text{Re},\phi_\text{Im},\phi'_\text{Re},\phi'_\text{Im})={1\over 2}\left[(\phi'_\text{Re})^2+(\phi'_\text{Im})^2+\mu (\phi^2_\text{Re}+\phi^2_\text{Im})\right] +{1\over 4}\Gamma(x) \left(\phi^2_\text{Re}+\phi^2_\text{Im}\right)^2$. Because $\Gamma(x) = \Gamma_{+} \; \chi_{[0,\infty)}(x) + \Gamma_{-} \; \chi_{(-\infty,0)}(x)$, the Hamiltonian $H$ is conserved on each $x\geq 0$ and $x<0$. As $\phi\in H^1(\R)$, both $\phi(x)$ and $\phi'(x)$ decay to $0$ as $|x|\rightarrow \infty$ so that $H(\phi_\text{Re}(x),\phi_\text{Im}(x),\phi'_\text{Re}(x),\phi'_\text{Im}(x))\rightarrow 0$ as $|x|\rightarrow \infty$, and thus $H\equiv 0$.

Differentiation yields 
\beq\label{E:H_deriv}
\begin{split}
0\equiv \frac{d}{dx}H =& \ \phi'_\text{Re} \left[\phi''_\text{Re} +\mu\phi_\text{Re} +\Gamma(x) \left(\phi^2_\text{Re} +\phi^2_\text{Im}\right) \phi_\text{Re} \right]+\\ 
& \ \phi'_\text{Im} \left[\phi''_\text{Im} +\mu\phi_\text{Im} +\Gamma(x) \left(\phi^2_\text{Im} +\phi^2_\text{Re}\right) \phi_\text{Im}\right]+\frac{1}{4}(\Gamma_+-\Gamma_-)\delta(x)\left(\phi^2_\text{Re} +\phi^2_\text{Im}\right)^2,
\end{split}
\eeq
where $\delta(x)$ is the Dirac delta.
Integrating \eqref{E:H_deriv} over $\R$, we get $0={1\over  4} (\Gamma_+-\Gamma_-)\left(\phi^2_\text{Re}(0) +\phi^2_\text{Im}(0)\right)^2$ so that $\phi_\text{Re}(0)=\phi_\text{Im}(0)=0$.

Finally, due to the $C^1$-nature of $\phi$ we have also $0=\lim_{x\rightarrow 0+}H(\phi_\text{Re}(x),\phi_\text{Im}(x),\phi'_\text{Re}(x),\phi'_\text{Im}(x))={1\over 2}\left((\phi'_\text{Re}(0))^2+(\phi'_\text{Im}(0))^2\right).$ In summary $\phi_\text{Re}(0)=\phi'_\text{Re}(0)=\phi_\text{Im}(0)=\phi'_\text{Im}(0)=0$. Hence $\phi \equiv 0$.
\end{proof}


As a result, solitons of the constant coefficient nonlinear Schr\"{o}dinger equation cannot be continued in $(\Gamma_--\Gamma_+)$ from $\Gamma_-=\Gamma_+$. This is due to the presence of the spatial shift invariance, which renders the corresponding Jacobian singular so that a parameter continuation fails. A periodic linear structure can thus be viewed as one of the simplest structures supporting localized waves at nonlinear interfaces.

In the rest of the paper we limit our attention to real SGSs profiles $\phi(x):\R\rightarrow \R$. These satisfy
\begin{align}\label{E:SPNLS}
\phi''(x) + \mu\phi(x) -  V(x) \phi(x) + \Gamma(x) \phi^3(x)=0, \quad x\in \R.
\end{align}
The $C^1$-solutions of \eqref{E:SPNLS} are critical points of the total energy 
\beq\label{E:energy}
E_{\mu}(\phi) = \int_{\mathbb{R}} \left[ (\phi')^2
- \mu \phi^2 + V(x) \phi^2 \right] dx - \frac{1}{2}
\int_{\mathbb{R}} \Gamma(x) \phi^4  dx,
\eeq
so that the first variation of $E_{\mu}(\phi)$ is equivalent to \eqref{E:SPNLS}.

Understanding stability of SGSs with respect to perturbations of the initial data $\phi(x)$ is of crucial importance for predicting physically and numerically observable solutions since experiments as well as numerical simulations generate unavoidable error. We investigate linear stability by inspecting the spectrum of the corresponding linearized operator. Standard methods for numerically approximating spectra of operators include direct eigenvalue computations in a finite dimensional approximation, e.g. finite differences or finite elements; Floquet-Bloch theory, where the problem is artificially treated as periodic \cite{Decon_Kutz06}; and the numerical Evans function method \cite{DG05,Brin01,PSK04,GW08}, out of which the last method is in principle restricted to one dimensional problems. Direct eigenvalue computations in a finite dimensional space typically suffer from spurious eigenvalues, which are difficult to identify \cite{DG05}. Although certain techniques for tackling this spectral pollution are available \cite{DaviesPlum04,Marletta09}, we choose the Evans function method, which does not suffer from spectral pollution. A further advantage of the Evans function method is the possibility to determine the number of eigenvalues within a region of the complex plane with a single contour integration  via the use of the argument principle. Nevertheless, we do not claim the superiority of this method over others. In the Evans function approach finding eigenvalues is reformulated as determining linear dependence of the stable and unstable manifolds of the zero solution of the linearized ODE. In the end this problem reduces to the evolution of an ODE system and the computation of a determinant, called the Evans function. Zeros of the Evans function coincide with eigenvalues of the linearized operator. As we show, a numerically stable and accurate evaluation of the Evans function, however, requires a careful computation of exponentially exploding solutions, the use of exterior algebra, which avoids numerical stiffness while preserving analyticity of the Evans function as well as the use of an ODE integrator which approximately preserves the weak Grassmanian invariant of the ODE system. Analyticity is essential in our method as we apply the argument principle in order to determine eigenvalues in the right half complex plane.

The aim of the paper is twofold. Firstly, it is to investigate properties of SGSs of \eqref{E:PNLS}, including their stability, and secondly to develop a practical numerical Evans function method combined with the argument principle for determining linear stability in problems with periodic coefficients. At the same time we point out possible difficulties of the method.

The rest of the paper is structured as follows. In the remaining part of the introduction we give a brief overview of Floquet theory and define some quantities needed in later sections. Section \ref{S:contin} discusses the numerical method of arclength continuation as applied to computing SGS families. Results of these computations are then presented and discussed for SGSs in the first two spectral gaps. Section \ref{S:stab} is devoted to linear stability of SGSs. After summarizing available information about the spectrum of the linearization we describe the Evans function method for locating the point spectrum. Next, we discuss difficulties in evaluating the Evans function in a numerically stable and accurate way and offer their solutions. Results of Evans function computations are then presented for the SGS families computed in Section \ref{S:contin}. Finally, direct numerical simulations of the PDE \eqref{E:PNLS} are performed for selected SGS examples to demonstrate their (un)stable evolution.

\subsection{Review of Floquet Theory}
The linear part of the stationary equation \eqref{E:SPNLS} is Hill's equation
\beq\label{E:Hill_sc}
Lq = -\pa_x^2q+V(x)q=\mu q, \quad x\in \R.
\eeq

An overview of Floquet theory for Hill's equation can be found in \cite{MagWin66,Eastham73}. The spectrum $\sigma(L)$ can be determined by studying the trace of the monodromy matrix of the first order version of \eqref{E:Hill_sc}. The spectrum is composed of bands, $\sigma(L) = \cup_{k\in \N}[s_{2k-1},s_{2k}]$, where $s_{2k+1}\geq s_{2k}>s_{2k-1}$. We define $\text{int}(\sigma(L)) = \cup_{k\in \N}(s_{2k-1},s_{2k})$ and $\pa \sigma(L) = \{s_1, s_2, \ldots \}$. For $\mu\in \C\setminus \pa \sigma(L)$ there are two linearly independent solutions of \eqref{E:Hill_sc}, so called Bloch waves, which have the form
\[q_1(x)=p_1(x) e^{\ri kx}, \quad q_2(x)=p_2(x) e^{-\ri kx}, \]
where $p_{1,2}(x+d)=p_{1,2}(x) \ \forall x\in \R$ and $k \in \C$. The periodic parts $p_{1,2}$ satisfy
\beq\label{E:shifted_evp}
-(\pa_x\pm\ri k)^2 p_{1,2}(x)+V(x)p_{1,2}(x) = \mu p_{1,2}(x), \quad x\in [0,d]
\eeq
with periodic boundary conditions.

For $\mu \in \text{int}(\sigma(L))$ the Bloch waves are bounded since $k\in \R$. For $\mu \in \pa \sigma(L)$ one solution is a periodic Bloch wave, and thus remains bounded, while the other linearly independent solution grows linearly in $x$.  We denote the periodic Bloch function at the edge $\mu =s_n \in \pa \sigma(L)$ by $q_n$. When $\mu\in \C\setminus \pa \sigma(L)$, the Bloch waves grow exponentially in either $x$ direction since $\text{Im}(k)\neq 0$.

\section{SGS Existence and Numerical Continuation}\label{S:contin}
In \cite{DP08} an implicit function theorem argument was used to show that gap solitons $\phi_{\text{GS}}$, i.e. solutions of \eqref{E:SPNLS} with $\Gamma\equiv \Gamma_0$, can be continued to SGSs as $\Gamma_-$ departs from $\Gamma_+=\Gamma_0$ provided the Jacobian $J = -\pa_x^2+V(x)-3\Gamma_0 \phi_{\text{GS}}^2(x)$ is nonsingular. This condition on $J$ is expected to hold due to the absence of shift invariance in equation \eqref{E:SPNLS}. Note that the continuation result holds for GSs centered at any $x_c \in \R$. In \cite{DP08} continuation in $\Gamma_-$ is carried out numerically for $x_c=0$ and $V(x) = \sin^2(\pi x/10)$ up to a critical value of $\Gamma_-$, where $J$ becomes singular. Here we show using the numerical arclength continuation method \cite{Keller77,Keller79} that the SGS families simply turn back with respect to $(\Gamma_--\Gamma_+)$ at this critical point and can be continued further. Often numerous turning points occur within an SGS family. These results suggest that the critical points are simple folds, where the kernel of $J$ is only one dimensional. Arclength continuation has been used in the context of the discrete NLS for instance in \cite{MFHKSFB06}.


Let us denote equation \eqref{E:SPNLS} as $G(\phi,\Gamma_-)=0$ with $\Gamma_+$ fixed and $\phi$ and $\Gamma_-$ unknown. Due to the jump in the coefficient $\Gamma$ the solution $\phi$ is only $C^1$ regular and we thus discretize \eqref{E:SPNLS} via $H^1$ conforming finite elements (using a FEM package by M. Richter from the Karlsruhe Institute of Technology). We use elements of 3rd order ($p=3$) on a domain $[-R,R]$ with $R>0$ large enough so that a given $\phi_{\text{GS}}$ is well decayed at $x=\pm R$ (typically $\phi_{\text{GS}}\approx 10^{-14}$ for $x\in [-R,-{9\over 10}R]\cup [{9\over 10}R,R]$). We use the element size $h\leq 0.06$ in all computations. As we show in Sec. \ref{S:evans_accur}, the accuracy of the Evans function, measured by how well the Evans function recovers the zero eigenvalue of the linearization, improves as $h \rightarrow 0$.

In order to describe the arclength continuation, let us denote the resulting FEM-discretized system of $N$ algebraic equations by
\beq\label{E:underdet}
\vec{G}(\vec{\phi},\Gamma_-)=0,
\eeq 
where $\vec{\phi} = (\phi_k)_{k=1}^N$ and $\phi_k$ is the approximation of $\phi(x_k), x_k = -R+(k-1)\frac{2R}{N-1}, \ k \in \{1,\ldots, N\}$. In the arclength continuation method the solution family $(\vec{\phi},\Gamma_-)$ is parametrized by the arclength $\tau$ of the curve generated by the family in the $(\|\vec{\phi}\|,\Gamma_-)$ plane with a selected norm $\|\cdot\|$, e.g. the $l_2$-norm. System \eqref{E:underdet} is underdetermined and the arclength method appends it with an equation that ensures continuation in the tangent direction. Namely, to extend the family from $\tau=\tau_0$ to $\tau=\tau_1$, one solves
\beq\label{E:arclength_eq}
\begin{split}
\vec{G}(\vec{\phi},\Gamma_-)&=0\\
\left(\vec{\phi}(\tau_1)-\vec{\phi}(\tau_0)\right)^T {d\over d\tau} \vec{\phi}(\tau_0)+\left(\Gamma_-(\tau_1)-\Gamma_-(\tau_0)\right){d\over d\tau}\Gamma_-(\tau_0)-(\tau_1-\tau_0)&=0
\end{split}
\eeq 
for the vector $(\vec{\phi}^T(\tau_1),\Gamma_-(\tau_1)) \in \R^{N+1}$, which we do by Newton's iteration using $(\vec{\phi}^T(\tau_0),\Gamma_-(\tau_0))+$$(\tau_1-\tau_0)({d\over d\tau} \vec{\phi}^T(\tau_0),{d\over d\tau} \Gamma_-(\tau_0))$ as the initial guess. After convergence the tangent vector at $\tau=\tau_1$ is obtained by solving 
\[\bpm \vec{G}_{\vec{\phi}}(\tau_1) & \vec{G}_{\Gamma_-}(\tau_1)\\
{d \over d\tau} \vec{\phi}(\tau_0) & {d \over d\tau} \Gamma_-(\tau_0)
\epm \bpm {d \over d\tau} \vec{\phi}(\tau_1) \\ {d \over d\tau} \Gamma_-(\tau_1)\epm = \bpm \vec{0} \\ 1\epm,\]
where $\vec{G}_{\vec{\phi}}$ is the $N\times N$ Jacobian matrix $\left({\pa \vec{G}\over \pa \phi_1}, \ldots, {\pa \vec{G}\over \pa \phi_N}\right)$ and $\vec{G}_{\Gamma_-}= {\pa \vec{G}\over \pa \Gamma_-}$.
The solution vector is normalized so that $\|{d \over d\tau} \vec{\phi}(\tau_1)\|_{l^2}^2+\left({d \over d\tau} \Gamma_-(\tau_1)\right)^2=1.$ The first continuation step from the GS at $\tau=0$, where the tangent vector is not available, is performed via a standard continuation in the parameter $\Gamma_-$. Newton's method is terminated when the $L^2$-norm of the residual decreases below a tolerance (typically $10^{-10}$). 

In \cite{PSK04} it is shown that families of GSs originate at upper edges of spectral gaps for the focusing nonlinearity $\Gamma_0>0$ and at lower gap edges for $\Gamma_0<0$. They bifurcate from the zero solution at the corresponding edge, which we call the \textit{bifurcation edge}.  For even potentials $V$ with one maximum and one minimum in each period there are two distinct families of GSs, namely those symmetric about the minimum location, so called onsite GSs, and those symmetric about the maximum location, offsite GSs \cite{PSK04}. GSs centered at $x=0$ can be computed via a numerical continuation in the parameter $\mu$ starting near a selected bifurcation gap edge $\mu=s_n$. As an initial guess for GSs near $s_n$ we use the slowly varying envelope approximation $\eps A_n(X)q_n(x), \ X=\eps x$, where $0<\eps^2<<1$ is the distance of $\mu$ to the gap edge: $\mu = s_n+\Omega \eps^2$ and $\Omega = \pm 1$ for lower and upper gap edges, respectively. $q_n$ is the periodic Bloch function at $\mu=s_n$ and $A_n$ satisfies the constant coefficient homogenized nonlinear Schr\"{o}dinger equation, see e.g. \cite{BSTU06}, Sec. IV in \cite{PSK04} or Sec. 3.3 in \cite{DP08},
\beq\label{E:NLS_amp}
\Omega A_n +\nu A''_n +\rho \Gamma A^3=0,
\eeq
where $\nu = 1+2(\tilde{q}_n',q_n)_{L^2(0,d)}$ and $\rho = \|q_n\|^4_{L^4(0,d)}$ with $\tilde{q}_n$ being a (periodic) `generalized Bloch function'. We use the explicitly known ground state solution of \eqref{E:NLS_amp}, namely $A_n(X)=\sqrt{{-2\Omega \over \rho\Gamma}}\sech\left(\sqrt{{-\Omega \over \nu}}X\right)$. In this sense we construct an approximation to `fundamental' GSs. Choosing an excited state solution to \eqref{E:NLS_amp}, a different family of GSs would result.

\subsection{Numerical Continuation Results}\label{S:num_cont}

We restrict our attention to the semi-infinite gap $(-\infty,s_1)$ and the first finite gap $(s_2,s_3)$ and first compute families of both on- and offsite GSs bifurcating from upper gap edges $s_1$ and $s_3$ for $\Gamma\equiv 1$ and from the lower edge $s_2$ for $\Gamma\equiv -1$. In all of the numerical computations we use 
\beq\label{E:V}
V(x) = \sin^2\left(\pi{x-x_0\over d}\right), \quad d=10,
\eeq
where $x_0 = 0$ for onsite and $x_0=d/2=5$ for offsite GSs. The shift $x_0$ is introduced in order to ensure that both onsite and offsite GSs are centered at $x=0$, the interface location for the subsequent SGS study. The first 10 edges of the spectrum $\sigma(L)$ for \eqref{E:V} have been computed using the trace of the monodromy matrix \cite{Eastham73} to be
\[(s_1, s_2, \ldots, s_{10}) \approx (0.2832, 0.2905, 0.7467, 0.8433, 1.0568, 1.4069, 1.4505, 2.0989, 2.1022, 2.9806).\]
Note that the edges $s_4, \ldots, s_{10}$ are needed below in Figs. \ref{F:Evans_fn_plot}. Figs. \ref{F:GS_on}(a) and \ref{F:GS_off}(a) show families of both onsite and offsite GSs $\phi_{\text{GS}}$ in the $(\mu,\|\phi_{\text{GS}}\|_{L^2}^2)$ plane. In optics $\|\phi\|_{L^2}^2$ is called the total power of $\phi$. For the labeled points $A_{\text{on}}-H_{\text{on}}$ and $A_{\text{off}}-H_{\text{off}}$ the GS profiles are also plotted. The $\mu$-values of the selected GSs are listed below.

\begin{tabular}{lllll|llllll}
$A_{\text{on}}$: & $\mu=0.278$ & \quad & $E_{\text{on}}$: & $\mu=0.297$ & \qquad & $A_{\text{off}}$: & $\mu=0.278$ & \quad & $E_{\text{off}}$: & $\mu=0.297$ \\
$B_{\text{on}}$: & $\mu=0.16$ & \quad & $F_{\text{on}}$: & $\mu=0.293$ & \qquad &$ B_{\text{off}}$: & $\mu=0.16$ & \quad & $F_{\text{off}}$: & $\mu=0.293$ \\
$C_{\text{on}}$: & $\mu=0.722$ & \quad & $G_{\text{on}}$: & $\mu=0.499$ & \qquad & $C_{\text{off}}$: & $\mu=0.722$ & \quad & $G_{\text{off}}$: & $\mu=0.499$ \\
$D_{\text{on}}$: & $\mu=0.502$ & \quad & $H_{\text{on}}$: & $\mu=0.72$ & \qquad & $D_{\text{off}}$: & $\mu=0.55$ & \quad & $H_{\text{off}}$: & $\mu=0.72$ \\
\end{tabular}

\begin{figure}
\begin{center}
\epsfig{figure = 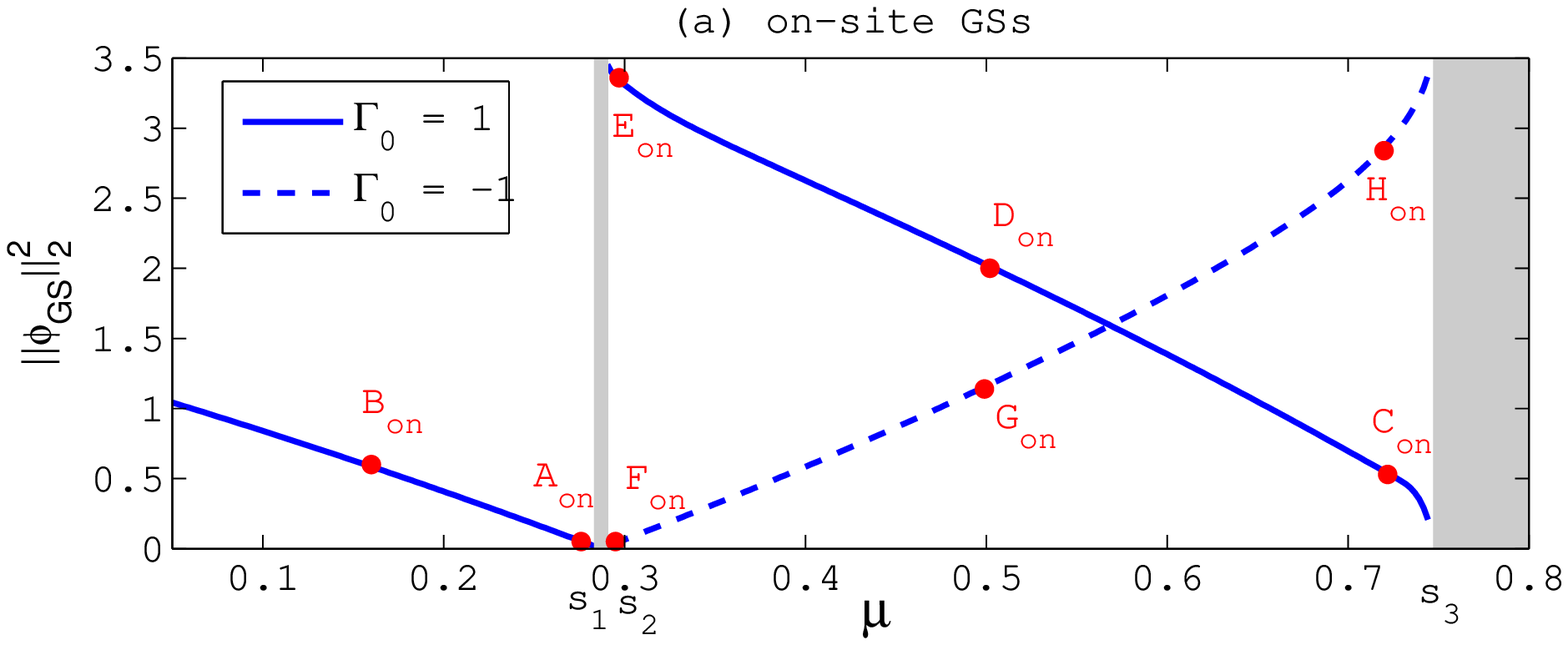,scale=.5}\\
\epsfig{figure = 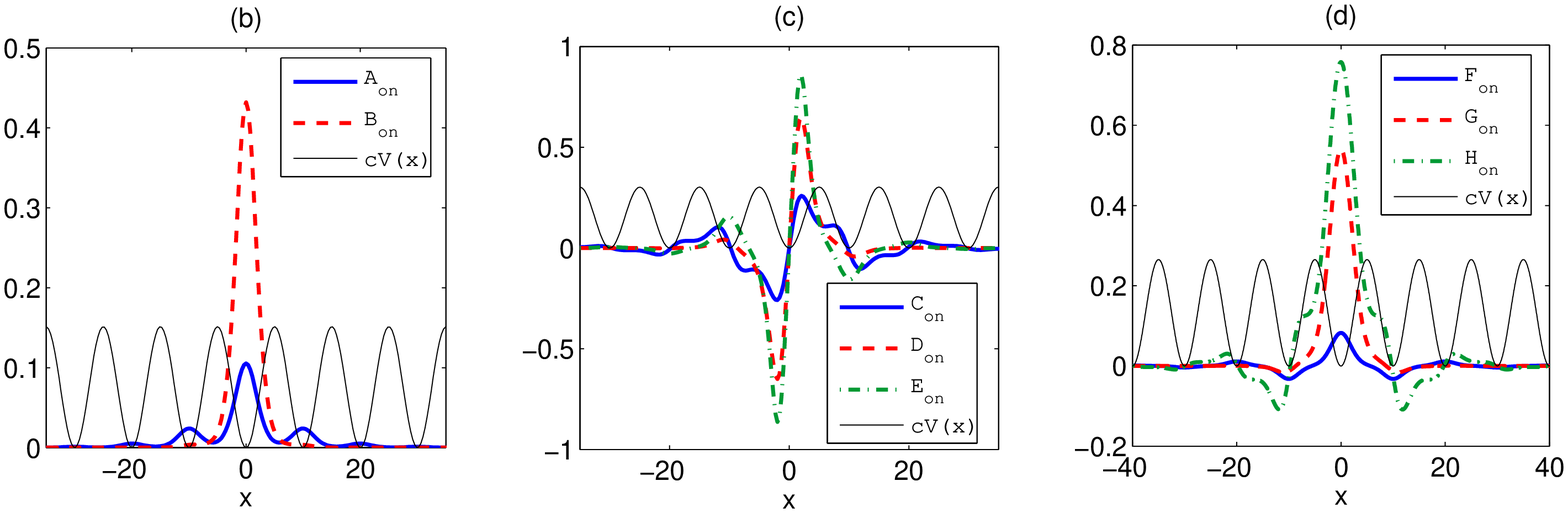,scale=.4}
\end{center}
\caption{Families of onsite gap solitons in the semi-infinite and the first finite gap for the potential \eqref{E:V} (with $x_0=0$). (a) GS continuation curves in the frequency -- total power coordinates; (b)-(d) GS profiles for the points labeled in (a). }
\label{F:GS_on}
\end{figure}
\begin{figure}
\begin{center}
\epsfig{figure = 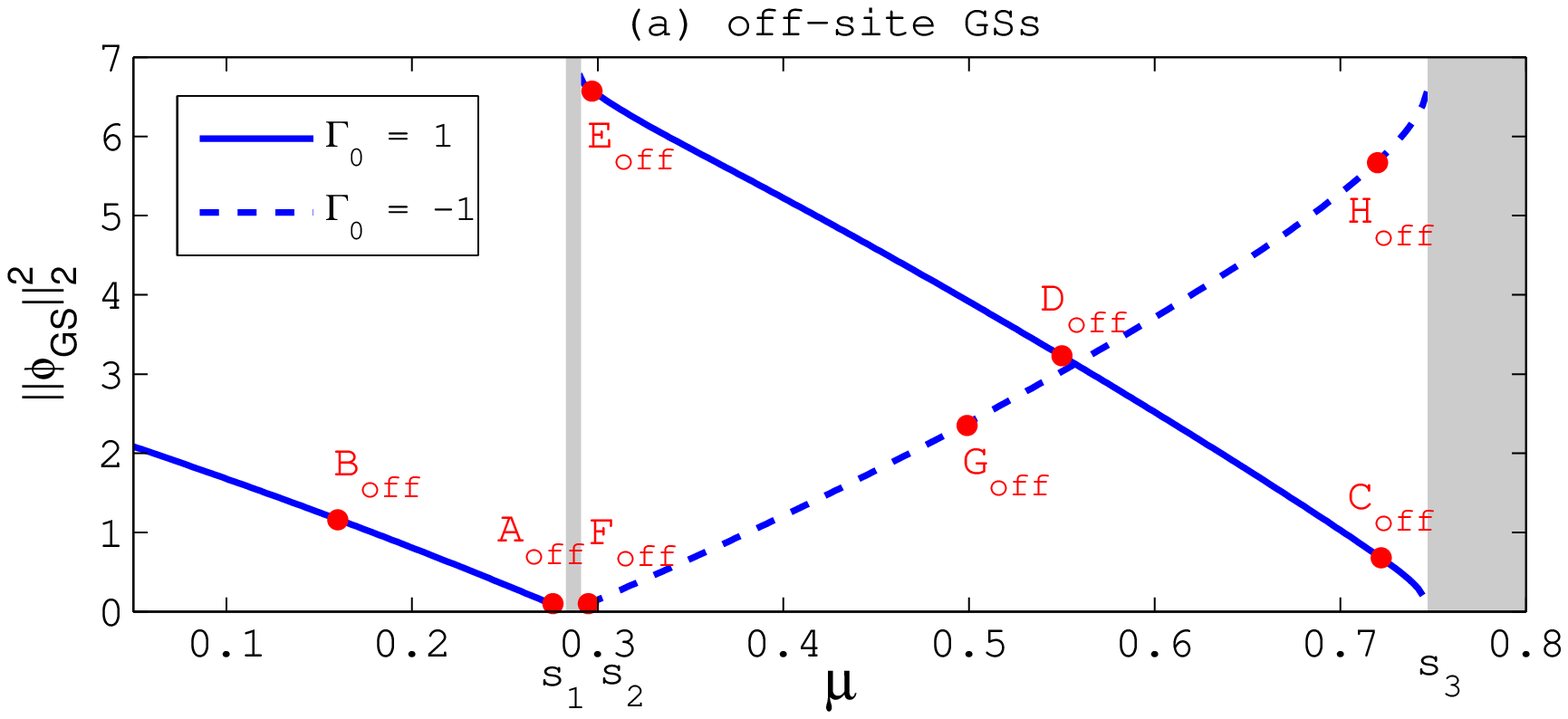,scale=.5}\\
\epsfig{figure = 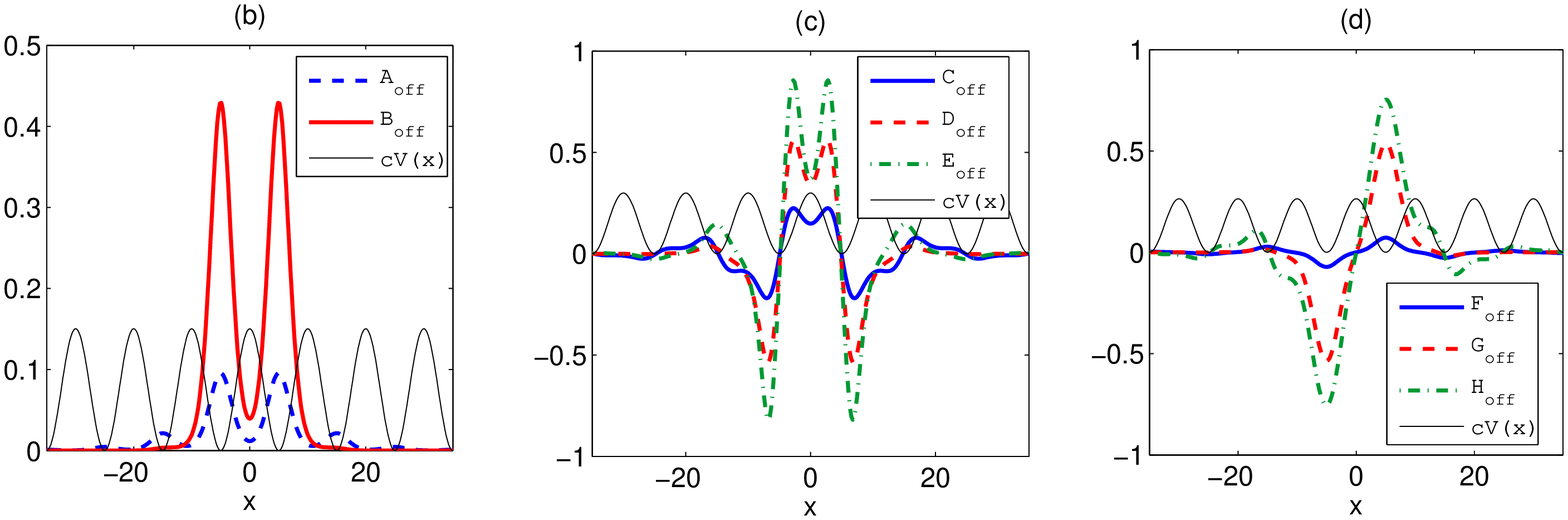,scale=.5}
\end{center}
\caption{Families of offsite gap solitons in the semi-infinite and the first finite gap for the potential \eqref{E:V} (with $x_0=d/2=5$). (a) GS continuation curves in the frequency -- total power coordinates; (b)-(d) GS profiles for the points labeled in (a).}
\label{F:GS_off}
\end{figure}

\begin{figure}
\begin{center}
\epsfig{figure = 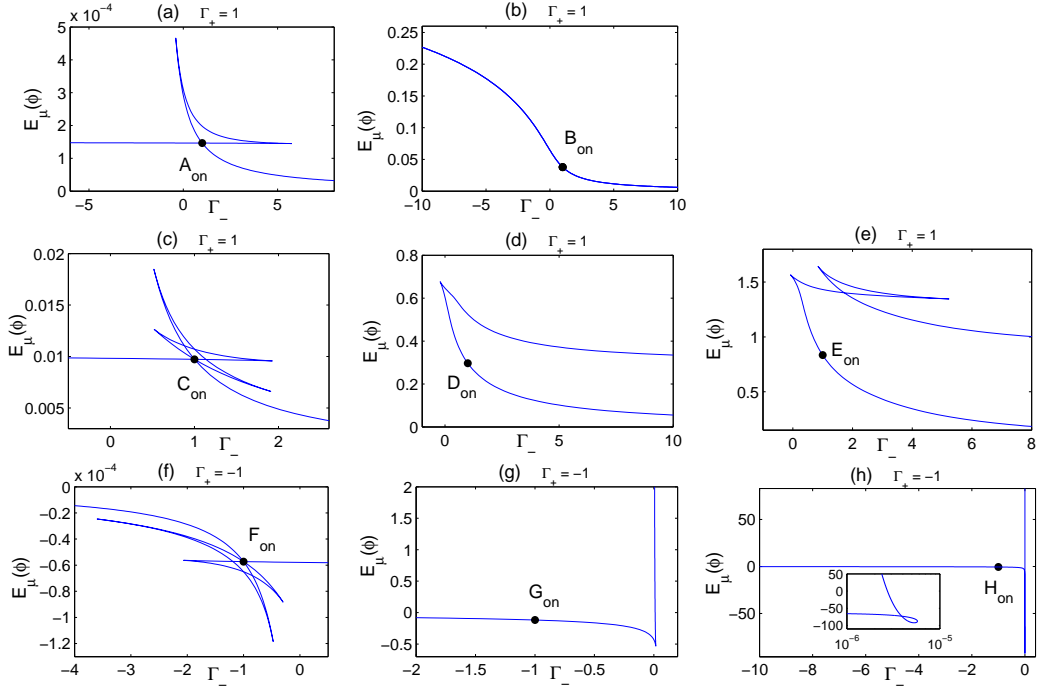,scale=.5}
\end{center}
\caption{SGS families bifurcating from the onsite GSs $A_{\text{on}}-H_{\text{on}}$ in Fig. \ref{F:GS_on}. Note that the curves were computed up to  $|\Gamma_-|=30$ or $|E_\mu(\phi)| = 100$, whichever occurred first,  and the behavior is qualitatively the same as in the displayed windows.}
\label{F:SGS_on}
\end{figure}

\begin{figure}
\begin{center}
\epsfig{figure = 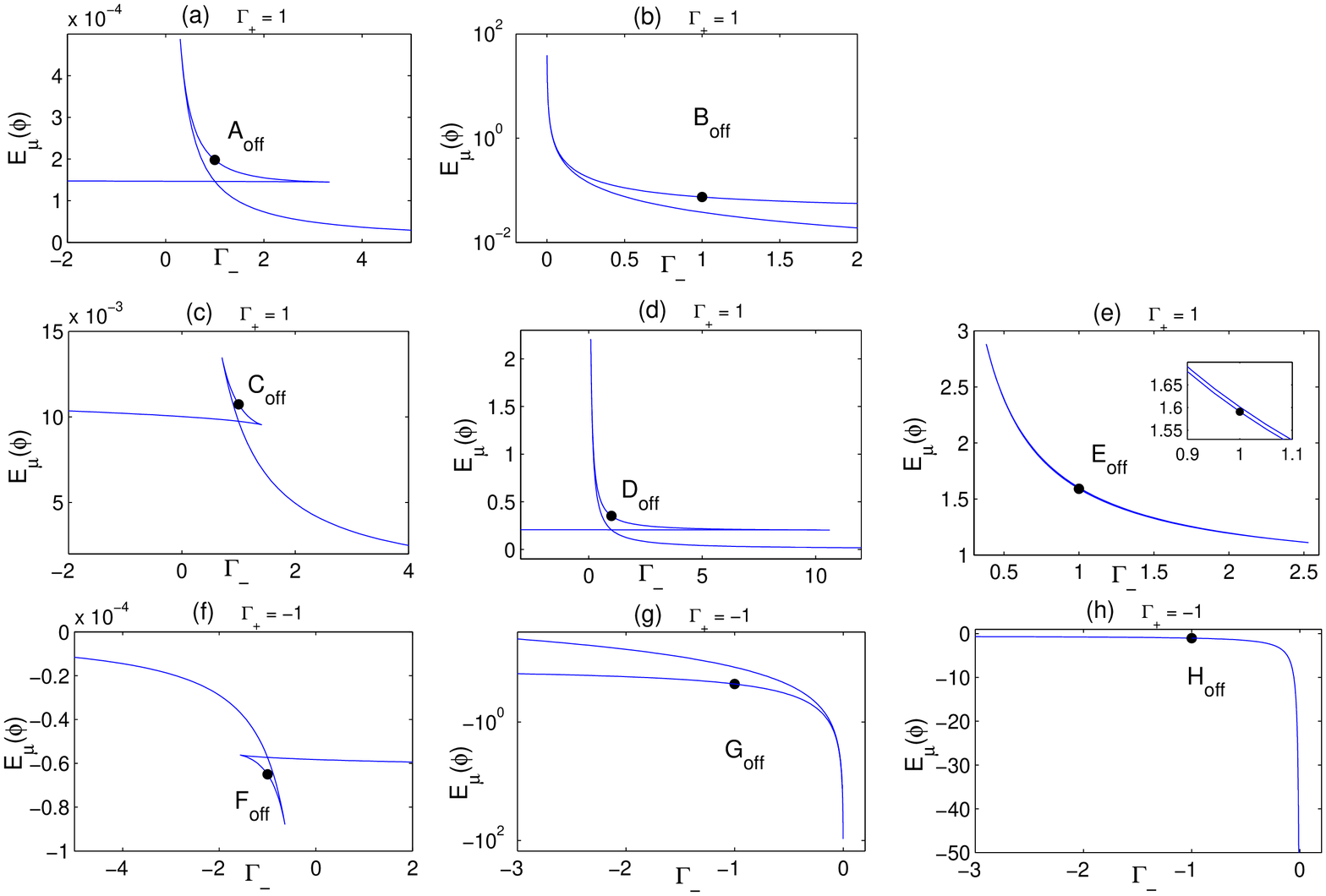,scale=.5}
\end{center}
\caption{SGS families bifurcating from the offsite GSs $A_{\text{off}}-H_{\text{off}}$ in Fig. \ref{F:GS_off}. Note that the curves were computed up to  $|\Gamma_-|=30$ or $|E_\mu(\phi)| = 100$, whichever occurred first,  and the behavior is qualitatively the same as in the displayed windows.}
\label{F:SGS_off}
\end{figure}

\begin{figure}
\begin{center}
\epsfig{figure = 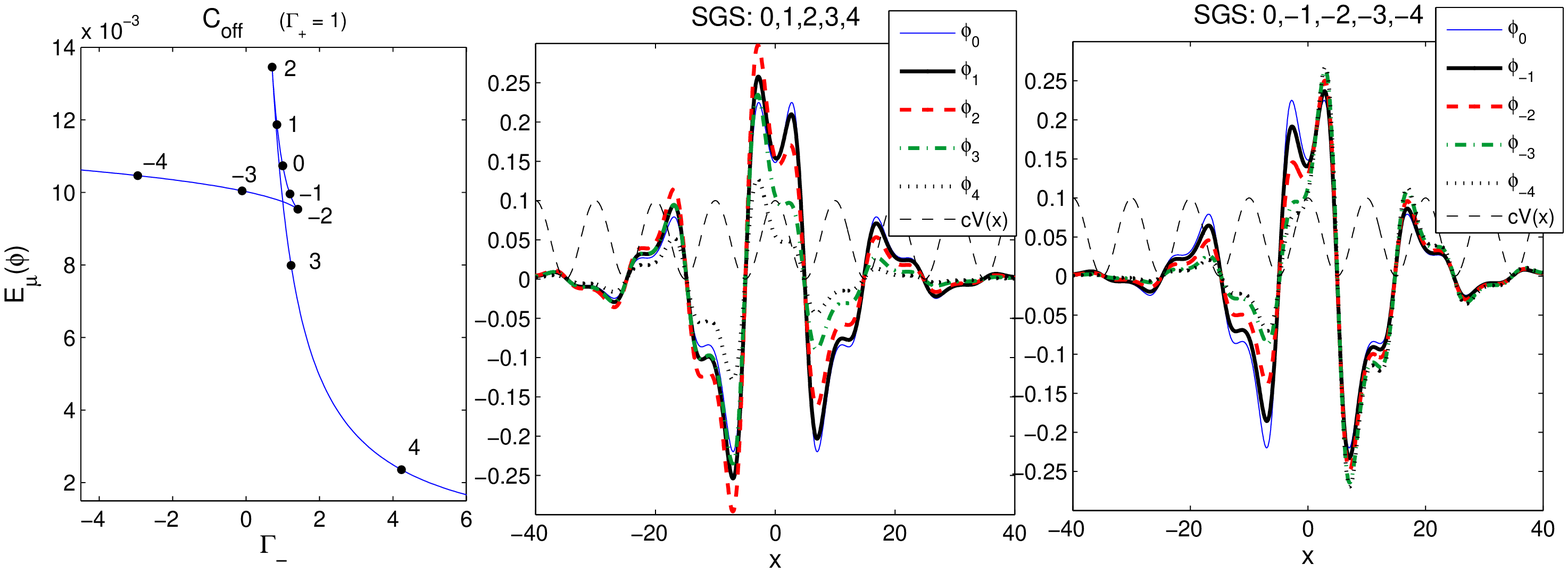,scale=.5}
\end{center}
\caption{Profiles of SGSs in the $C_{\text{off}}$ family. (a) the SGS family in the $(\Gamma_-,E_\mu(\phi))$ plane with 9 points labeled; (b) and (c) profiles of the SGSs labeled in (a).}
\label{F:C_off_SGS_profs}
\end{figure}

Families of SGSs bifurcating from these GSs have been computed next via the arclength continuation described above holding $\Gamma_+$ fixed at the corresponding GS value, i.e., either $\Gamma_+=1$ or $\Gamma_+=-1$.  The points $A_{\text{on}}-H_{\text{on}}$ and $A_{\text{off}}-H_{\text{off}}$ have been selected as representative examples. This set includes points near bifurcation edges, where $\|\phi\|_{L^2}\rightarrow 0$ as $\mu$ approaches the edge, near the opposite edges, and also deep inside the gaps. Figs. \ref{F:SGS_on} and \ref{F:SGS_off} show these SGS families in the $(\Gamma_-,E_\mu(\phi))$ plane for the onsite and offsite case, respectively. The choice of $E_\mu(\phi)$ over $\|\phi\|_{L^2}^2$ has been made due to a clearer identification of fold locations in the former case. The folds are shown in Sec. \ref{S:stab_numerics} to often correspond to stability changes of the SGS and in all the considered cases they are accompanied with a change in the number of eigenvalues in the right half complex plane. The black dots in Figs.  \ref{F:SGS_on} and \ref{F:SGS_off} correspond to the GSs from which continuation was started in both the $\Gamma_-<\Gamma_+$ and $\Gamma_->\Gamma_+$ directions. In Fig. \ref{F:C_off_SGS_profs} we plot for illustration the profiles of several SGSs along the family $C_\text{off}$. 

\subsubsection{Observations on Numerical Results}



\paragraph{Occurrence of Translated Bifurcation GSs}

\begin{figure}
\begin{center}
\epsfig{figure = 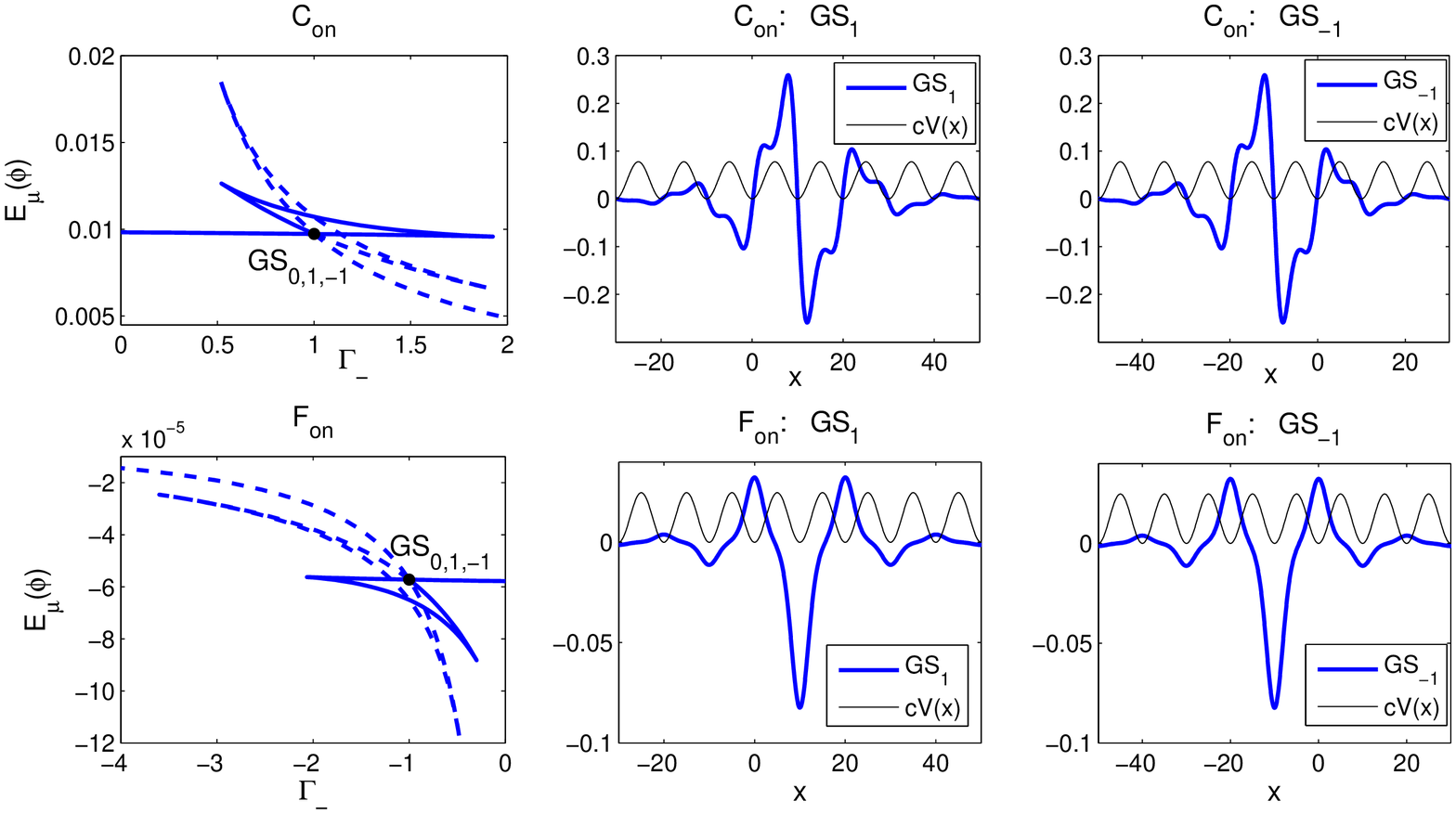,scale=.5}
\end{center}
\caption{Translated bifurcation GSs within SGS families. Left: Families of SGSs in the case $C_{\text{on}}$ and $F_{\text{on}}$ at the top and bottom resp. Middle: GS at the (first) return of the full line in the left plot to $(\Gamma_+,E_\mu(\phi_{\text{GS}}))$. These GSs, denoted GS$_1$ are centered at $x=d$. Right: GS at the (first) return of the dashed line in the left plot to $(\Gamma_+,E_\mu(\phi_{\text{GS}}))$. These GSs, denoted GS$_{-1}$, are centered at $x=-d$.}
\label{F:GS_return}
\end{figure}

As one can see in Fig. \ref{F:SGS_on} (a), (c), and (f), families of SGSs may return to the point corresponding to the bifurcation GS, i.e. $(\Gamma_-,E_\mu(\phi))=(\Gamma_+,E_\mu(\phi_{\text{GS}}))$. In fact, we show in Fig. \ref{F:GS_return} (using the families $C_{\text{on}}$ and $F_{\text{on}}$ as examples) that the solutions corresponding to all these intersections of $(\Gamma_+,E_\mu(\phi_{\text{GS}}))$ are GSs. Each return to $(\Gamma_+,E_\mu(\phi_{\text{GS}}))$ is, however, a GS centered at a different extremum of $V$. We expect that generically for a bifurcation from an onsite GS $\phi_{\text{GS}}$ centered at $x=0$ each half of the SGS family, i.e., the one bifurcating in $\Gamma_-$ to the left of $\Gamma_+$ and the one bifurcating to the right, the $n-$th return to $(\Gamma_+,E_\mu(\phi_{\text{GS}}))$ is an onsite GS centered at $x=nd$ or $x=-nd$. This has been checked to hold in all the computed examples with SGS families that are open curves in the  $(\Gamma_-,E_\mu(\phi))$ plane. Note that it is impossible to conclude based on the equalities $\Gamma_-=\Gamma_+$ and $E_\mu(\phi)=E_\mu(\phi_{\text{GS}})$ that the solution $\phi$ \textit{must} be a GS but our numerics suggest this. Fig. \ref{F:GS_return} demonstrates this for the cases $C_{\text{on}}$ and $F_{\text{on}}$, where returns with $n=1$ occur for both. 

\begin{figure}
\begin{center}
\epsfig{figure = 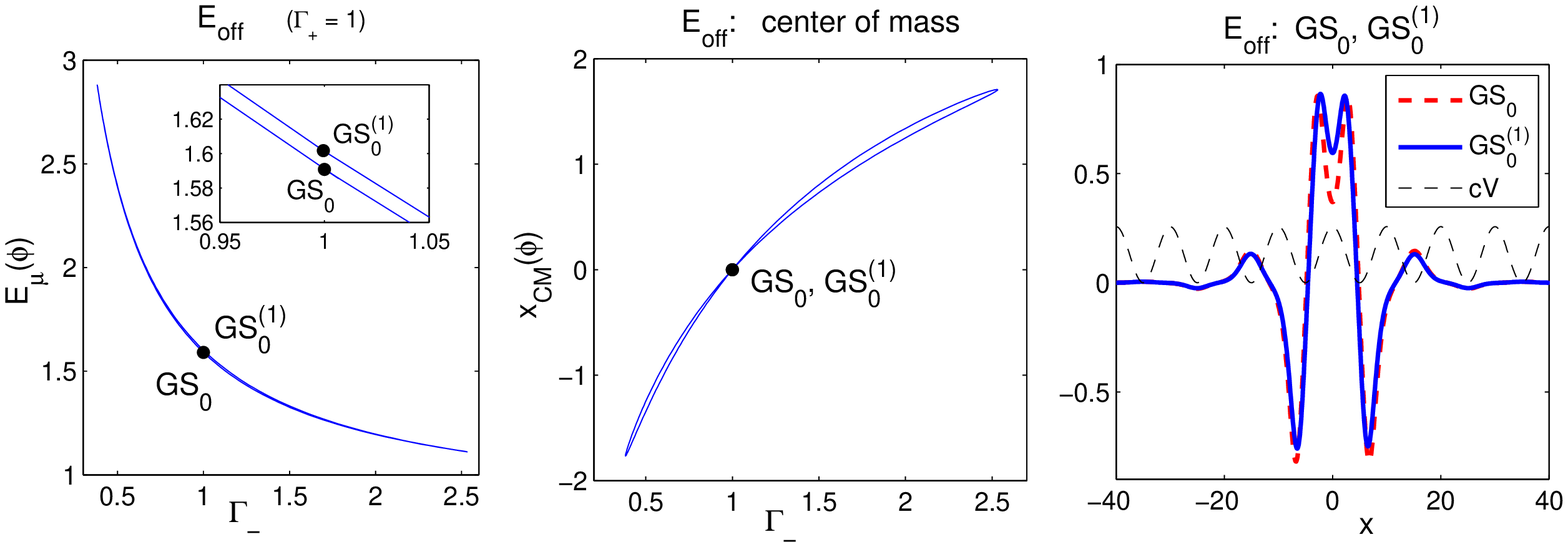,scale=.5}
\end{center}
\caption{Two GSs centered at $x=0$ within the $E_{\text{off}}$ family. (a) $(\Gamma_-,E_\mu(\phi))$ curve of the SGS family; (b) center of mass; (c) profiles of the two solutions on the $\Gamma_-=\Gamma_+$ line. Both are GSs centered at $x=0$.}
\label{F:E_off_nonunique_GS}
\end{figure}

For closed SGS family curves, like the case $E_{\text{off}}$ in Fig. \ref{F:SGS_off} this rule is not expected to hold. Upon return of the family $E_{\text{off}}$ to $(\Gamma_-,E_\mu(\phi))=(\Gamma_+,E_\mu(\phi_{\text{GS}}))$ the solution is, indeed, identical to the bifurcation GS $\phi_{\text{GS}}$ centered at $x=0$. As Fig. \ref{F:E_off_nonunique_GS} shows, the other solution that lies on the line $\Gamma_-=\Gamma_+$ and belongs to the $E_{\text{off}}$ family is also an offsite GS centered at $x=0$ but different from $\phi_{\text{GS}}$. This is possible due to non-uniqueness of bound states of \eqref{E:SPNLS}. Part (b) of Fig. \ref{F:E_off_nonunique_GS} plots the center of ``mass'' 
\[x_{CM}(\phi):=\int_\R x|\phi(x)|dx \left / \int_\R |\phi(x)|dx \right. \]
for the family and shows that the whole family $E_{\text{off}}$ consists of solutions whose center of mass does not move far from the minimum point $x=0$ of $V$ compared, e.g., to the cases $C_{\text{on}}, F_{\text{off}}$ below.

It is an open question how many returns to $(\Gamma_-,E_\mu(\phi))=(\Gamma_+,E_\mu(\phi_{\text{GS}}))$ occur in a given SGS family.
Note that none of the open families of SGSs in Fig. \ref{F:SGS_off}, which bifurcate from offsite GSs, returns to $(\Gamma_+,E_\mu(\phi_{\text{GS}}))$ in contrast with the onsite case where these returns occur, e.g. for $A_\text{on}, C_\text{on},$ and $F_\text{on}$, see Fig. \ref{F:SGS_on}. Nevertheless, we have found offsite cases where these returns do occur in the second finite gap, e.g. at $\mu \approx 0.861$ and $\Gamma_+=1$.

\paragraph{Homotopy between Onsite and Offsite GSs.}

\begin{figure}
\begin{center}
\epsfig{figure = 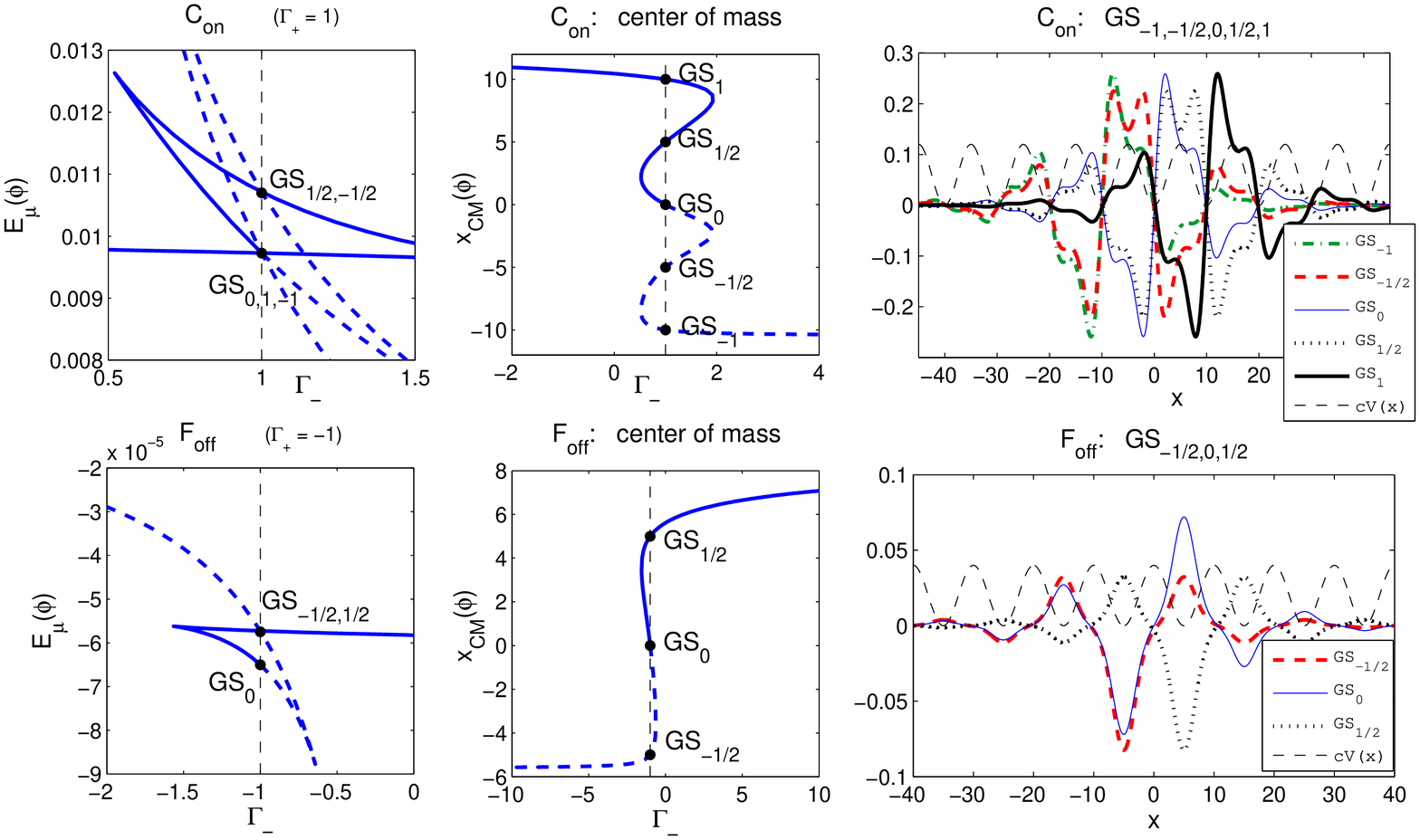,scale=.5}
\end{center}
\caption{On- and offsite GSs within an SGS family. Top and bottom: SGS families for the case $C_{\text{on}}$ and $F_{\text{off}}$ resp. Left: total energy $E_\mu(\phi)$; middle: center of mass $x_\text{CM}(\phi)$; right: profiles of GSs marked on the left and in the middle. For the case $C_\text{on}$ (where $x_0=0$) profiles GS$_n$ with integer indices $n\in \Z$ are onsite GSs while those with indices $n/2, n \in \Z$ are offsite GSs. For $F_\text{off}$ (where $x_0=5$) this is vice versa. The dashed vertical lines correspond to a no interface medium ($\Gamma_-=\Gamma_+$).}
\label{F:SGS_vert_line}
\end{figure}

As one can see in Fig. \ref{F:SGS_on} and \ref{F:SGS_off}, SGS families often intersect the vertical line $\Gamma_-=\Gamma_+$, which describes a no-interface medium. The numerical results show that at each of these intersections the corresponding solution $\phi$ is a symmetric GS. Clearly, for \eqref{E:V} with $x_0=0$ (used in SGS bifurcations from onsite GSs) if a GS is centered at $x=nd, \ n\in \Z$, it is an onsite GS and if it is centered at $x=(2n+1){d\over 2}, \ n \in \Z$, it is an offsite GS. In the case $x_0={d\over 2}$ (used in SGS bifurcations from offsite GSs) if the center of a GS is at $x=nd$ or $x=(2n+1){d\over 2}$, it is an offsite or onsite GS respectively. Fig. \ref{F:SGS_vert_line} plots the GS solutions located on the vertical line $\Gamma_-=\Gamma_+$ for the families $C_{\text{on}}$ and $F_{\text{off}}$. It also shows the center of mass for all the SGSs along the families clearly visualizing that the SGS is no always shifted to the more focusing medium. 

\begin{figure}
\begin{center}
\epsfig{figure = 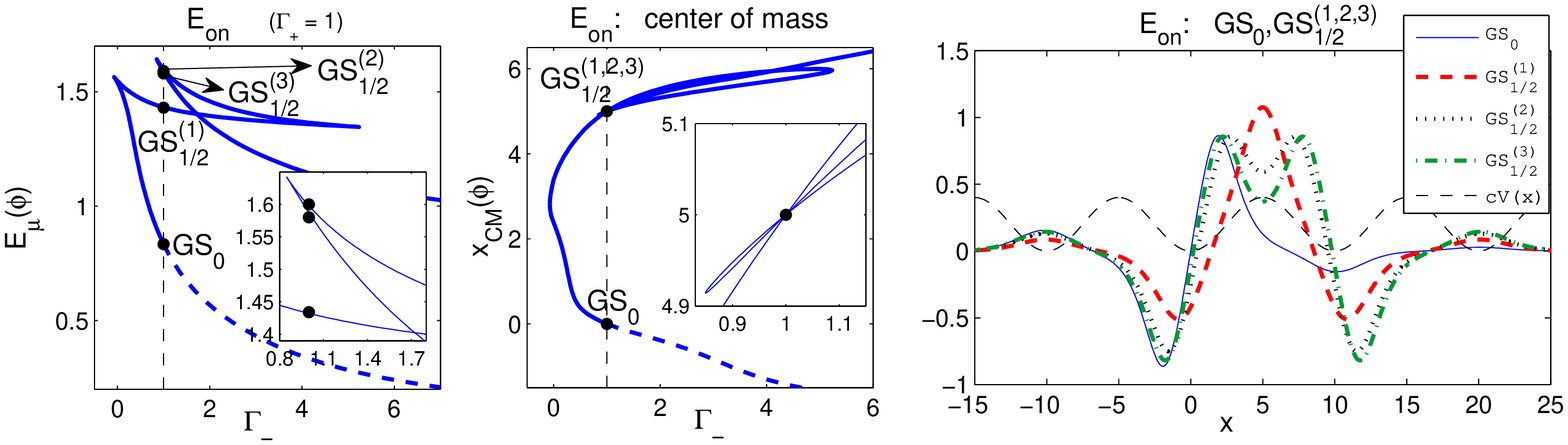,scale=.5}
\end{center}
\caption{SGS family through $E_\text{on}$; description as in Fig. \ref{F:SGS_vert_line}. Notice the non-monotonous dependence of the center of mass on the arclength in contrast to the cases in Fig. \ref{F:SGS_vert_line}.}
\label{F:SGS_vert_line_loop}
\end{figure}

Although in Fig. \ref{F:SGS_vert_line} the function $x_{CM}(\phi)$ appears to be monotonic in the arclength $\tau$, this is not always the case; not even for open SGS families. Fig. \ref{F:SGS_vert_line_loop} shows the family $E_{\text{on}}$ from Fig. \ref{F:SGS_on} (e) for which four intersections of $\Gamma_-=\Gamma_+$ occur but three of these are GSs centered at $x=d/2=5$. Note also that these three GSs are all different. 

\begin{figure}
\begin{center}
\epsfig{figure = 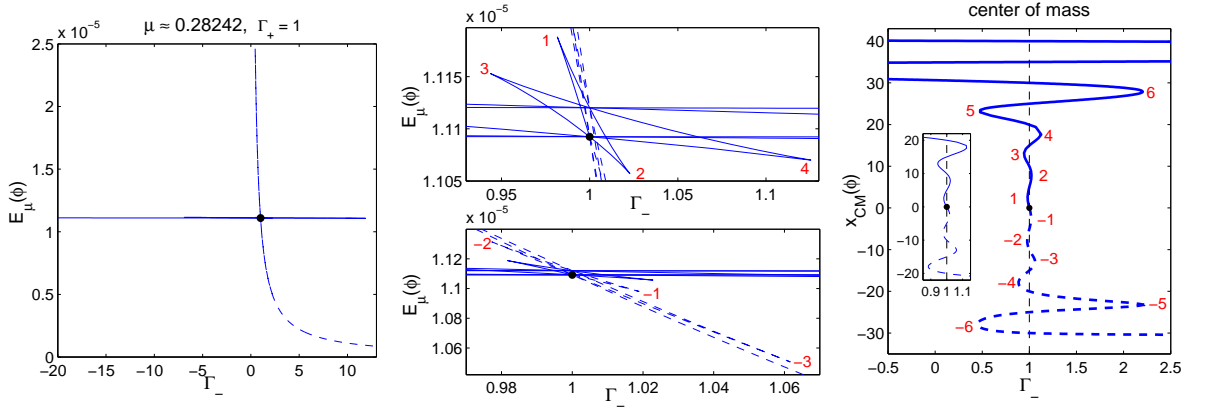,scale=.5}
\end{center}
\caption{Behavior near gap edges. SGS family bifurcation from the onsite GS at $\mu \approx 0.28242$ (near the minimum of the first band) with $\Gamma_+=1$. Left: total energy $E_\mu(\phi)$; middle: zoomed-in parts of the plot on the left; right: center of mass. Notice the large number of folds near the GS bifurcation point $(\Gamma_-,E_\mu(\phi))=(\Gamma_+,E_\mu(\phi_\text{GS}))$.}
\label{F:SGS_fam_near_edge}
\end{figure}

The numerics suggest that each shift of the SGS center from one extremum of $V$ to another is accompanied by a fold. Moreover, for spatially relatively broad SGSs, i.e., when $\mu$ lies near the corresponding bifurcation edge, all folds seem to correspond to such shifts, see Fig. \ref{F:SGS_vert_line} and \ref{F:SGS_fam_near_edge}. The same occurs also for the cases $A_\text{on},F_\text{on},A_\text{off}$ and $C_\text{off}$.

\paragraph{Asymptotics near Bifurcation Gap Edges.}

It was shown in \cite{DP08} that in the vicinity of spectral gap edges $\mu =s_n$ the interval of existence $\Gamma_- \in (\Gamma_+-a(\mu),\Gamma_++b(\mu))$ with  $a(\mu),b(\mu)>0$ for SGSs localized near $x=0$ collapses as $\mu \rightarrow s_n$. In detail $a(\mu), b(\mu) \rightarrow 0+$ as $\mu \rightarrow s_n-$ for $\Gamma_+>0$ and as $\mu \rightarrow s_n+$ for $\Gamma_+<0$. In practice, since folds typically correspond to transitions of SGS centers to the next extremum location of $V$, this means that the horizontal distance $|\Gamma_- - \Gamma_+|$ of any GS point along the SGS family to the nearest fold converges to $0$ as the edge is approached. Here we show that the same holds for SGSs localized around any point $x_c\in \R$. Analogously to Sec. 3.3 in \cite{DP08} we use the asymptotic representation 
\beq\label{E:asympt_edge}
\begin{split}
\mu & = s_n + \eps^2 \Omega + O(\eps^4)\\
\phi(x) &= \eps A(X)q_n(x) + \eps^2 A'(X) \tilde{q}_n(x)+\eps^3\phi^{(3)}(x,X) + O(\eps^4)
\end{split}
\eeq
with $X=\eps(x-x_c), \ 0<\eps<<1$. Substituting this two-scale expansion in \eqref{E:SPNLS}, we get at $O(\eps)$ and $O(\eps^2)$
\beq\label{E:asymp_O12}
(L-s_n)q_n=0 \qquad \text{and} \qquad (L-s_n)\tilde{q}_n=2q_n'
\eeq
respectively, which both have periodic solutions. At $O(\eps^3)$ we have
\beq\label{E:phi3_eq}
(L-s_n)\phi^{(3)} = \Omega A_n q_n + A''(q_n+2\tilde{q}'_n)+\Gamma(X+\eps x_c) A^3 q_n^3,
\eeq
where we have used the fact that due to the form $\Gamma(x) = \Gamma_{+} \; \chi_{[0,\infty)}(x) + \Gamma_{-} \; \chi_{(-\infty,0)}(x)$ it is
$\Gamma(x)=\Gamma(\eps x)=\Gamma(X+\eps x_c)$.

To ensure the existence of a periodic solution $\phi^{(3)}$, the Fredholm alternative needs to be applied to \eqref{E:phi3_eq} resulting in
\beq\label{E:NLS_edge}
\Omega A +\nu A'' +\rho \Gamma(X+\eps x_c) A^3=0
\eeq
with $\nu$ and $\mu$ defined in \eqref{E:NLS_amp}.

Due to Lemma \ref{L:nonexistence_homog} equation \eqref{E:NLS_edge} has non-trivial localized solutions only for $\Gamma_+=\Gamma_-$. For the SGS families this means that as $\mu$ approaches the corresponding spectral edge $s_n$, they lie inside narrower and narrower vertical slabs $\Gamma_-\in (\Gamma_+-a(\mu),\Gamma_++b(\mu))$ with $a(\mu),b(\mu) \rightarrow 0$ as $\mu$ approaches $s_n$. In other words the horizontal distance $|\Gamma_- - \Gamma_+|$ of \textit{every} fold to the GS bifurcation point converges to $0$ as $\mu$ approaches $s_n$. The convergence rate is, of course, expected to decrease with $|n|$, where $n$ is the signed ordinal of the fold with respect of the bifurcation GS, since the effect of the interface decreases with increasing distance between the solution center and the interface location $x=0$. In fact, one can conjecture an exponentially fast decrease of this rate as $|n|$ increases due to the exponential decay rate of SGS tails. In other words, we expect that for a fixed $\mu$ with $|\mu-s_n|<<1$ the distance of the $n$-th fold grows exponentially with $|n|$, cf. the right panel of Figure~\ref{F:SGS_fam_near_edge}.

For GSs centered at $x=0$ the first fold in each $\Gamma_-$ direction was followed numerically in $\mu$ in \cite{DP08}. Although we have not followed other folds systematically, our numerical results are in agreement with this asymptotic statement. Fig. \ref{F:SGS_fam_near_edge} presents a part of an SGS family bifurcating from the onsite GS at $\mu \approx 0.28242$, i.e., close to the edge $s_1 \approx 0.28317$. The first 4 folds in both directions $|n|\leq 4$ occur for $|\Gamma_--\Gamma_+|<0.13$. The part of the family between the folds $n=-4$ and $n=4$ consists of solutions with centers of mass in $(-20,20)$. It thus contains also solutions centered relatively far from the origin. 
For $\mu \approx 0.28278$, i.e., even closer to $s_1$, solutions between the first 18 folds ($|n|\leq 9$) have $|\Gamma_--\Gamma_+|<0.1$ and centers of mass in $(-50,50)$. The plot of the curve $(\Gamma_-,E_\mu(\phi))$ is even more complicated and less readable than in Fig. \ref{F:SGS_fam_near_edge}.


\section{Linear Stability of SGSs}\label{S:stab}

One of the most important properties of the above computed SGSs of \eqref{E:PNLS} is their stability with respect to perturbations of initial data. Only stable SGSs can be viewed as physically relevant states of the system. Orbital stability of bound states $e^{-\ri \mu_s t}\phi_s(x)$ of \eqref{E:PNLS} with $\phi_s$ positive and $\mu_s$ in the semi-infinite gap can be checked using the Vakhitov-Kolokolov criterion on the sign of ${dP(\phi(\cdot,\mu_s))\over d\mu}$, where $P(\phi(\cdot,\mu_s))=\|\phi(\cdot,\mu_s)\|_{L^2(\R)}^2$, and on the kernel and the negative part of the spectrum of the operator $L_+$ (defined below in the proof of Lemma \ref{L:bound_sp}) \cite{W86,GSS87,SFIW08}. In detail, orbital stability is satisfied if ${dP(\phi(\cdot,\mu_s))\over d\mu}<0$ and $L_+$ has no zero eigenvalues and the number of its negative eigenvalues is $1$. These conditions have to be in general checked numerically and, in particular, the spectral condition on $L_+$ is not completely trivial to check. Moreover, this test applies only in the semi-infinite gap. In order to be able to make stability statements even about SGSs in finite gaps, we choose to study \textit{linear stability} by inspecting directly the 
spectrum of the linearized operator via the \textit{numerical Evans function method}.

Given a SGS $e^{-\ri\mu_s t}\phi_s(x)$, we consider the evolution of a perturbed solution $\psi(x,t)=e^{-\ri \mu_s t}(\phi_s(x)+q(x,t))$ with $q(x,0)$ small. Linearizing \eqref{E:PNLS} in $q$ and using that $(\mu_s,\phi_s)$ satisfies \eqref{E:SPNLS}, we obtain
\begin{align}
\ri \partial_t q + \mu_s q + \partial_x^2 q - V(x) q + \Gamma(x) \phi_s^2(x)(\overline{q} + 2 q) = 0. \label{E:linearized}
\end{align}
As a system for $(U_1,U_2):=(q,\bar{q})$ this becomes
\begin{equation}\label{E:spectral_ODE}
\partial_t \begin{pmatrix} U_1 \\ U_2 \end{pmatrix} = \mathbb{L} \begin{pmatrix} U_1 \\ U_2 \end{pmatrix} 
\end{equation}
with
\begin{equation}
\mathbb{L} := -\ri \begin{pmatrix} -\partial_x^2 -\mu_s + V(x) - 2 \Gamma(x) \phi_s^2 & -\Gamma(x) \phi_s^2 \\ \Gamma(x) \phi_s^2 & \partial_x^2 +\mu_s - V(x) + 2 \Gamma(x) \phi_s^2 \end{pmatrix} 
\label{E:linearizedOp}
\end{equation}
As \eqref{E:spectral_ODE} is $t-$autonomous, the problem is separable and we can let $(U_1(x,t),U_2(x,t))=e^{\lambda t}(u_1(x),u_2(x))$ so that we obtain the spectral problem 
\beq\label{E:evp}
\lambda \begin{pmatrix} u_1 \\ u_2 \end{pmatrix} = \mathbb{L} \begin{pmatrix} u_1 \\ u_2 \end{pmatrix}.
\eeq
Clearly, linear stability of $e^{-\ri\mu_s t}\phi_s(x)$ is determined by the real part of $\sigma(\L)$.

\subsection{The Spectrum of $\L$}

\begin{figure}
\begin{center}
\epsfig{figure = 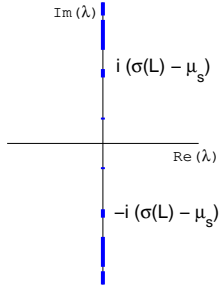,scale=.5}
\end{center}
\caption{The essential spectrum of $\L$ for $V(x)=\sin^2(\pi x/10)$.}
\label{F:ess_sp}
\end{figure}

Although $\sigma(\L)$ consists of the essential spectrum $\sigma_{\text{ess}}(\L)$ and of eigenvalues $\sigma_{\text{disc}}(\L)$, only eigenvalues dictate stability properties in this case because $\sigma_\text{ess}(\L)\subset \ri \R$. This follows from Weyl's theorem, see Sec. XIII.4 of \cite{ReedSimon4}, which ensures that due to the exponential decay of $\phi_s$, the terms containing $\phi_s$ are a relatively compact perturbation of $\L$ and the essential spectrum of $\L$ equals that of $\ri \left(\begin{smallmatrix}L-\mu_s& 0\\ 0&-(L-\mu_s)\end{smallmatrix}\right)$. As this operator is skew-adjoint, its spectrum is imaginary. Moreover, we also have that 
\[\sigma_{\text{ess}}(\L) = \ri (\sigma(L)-\mu_s) \cup -\ri (\sigma(L)-\mu_s),\] 
where the first bands of $\sigma(L)$ are detailed at the beginning of Section \ref{S:num_cont}. Fig. \ref{F:ess_sp} plots  $\sigma_{\text{ess}}(\L)$ for the potential \eqref{E:V} schematically.

Although $\sigma_{\text{disc}}(\L)$ will be determined numerically, the following Lemma gives an upper bound on $\text{Re}(\sigma_{\text{disc}}(\L))$. For the dynamics of \eqref{E:spectral_ODE} this is an upper bound on the growth rate of the perturbation. Note, however, that the numerical method for studying $\sigma_{\text{disc}}(\L)$ will not make use of this result.
\begin{lemma}\label{L:bound_sp}
Eigenvalues $\lambda \in \sigma_{disc}(\mathbb{L})$ satisfy $|\text{Re}(\lambda)| \leq ||\Gamma||_{\infty} ||\phi_s||_{\infty}^2.$
\end{lemma} 
\begin{proof}
With the new set of variables $a := u_1+u_2$ and $b := -\ri(u_1-u_2)$ equation \eqref{E:evp} becomes
\beq\label{E:evp_re}
\lambda \bpm a \\b\epm = \bpm 0 &L_-\\
-L_+ & 0
\epm
\bpm a \\b\epm \qquad \text{where} \quad \begin{array}{ll}L_-&=-\pa_x^2 +V(x)-\mu_s -\Gamma(x) \phi_s^2(x),\\
L_+ &= -\pa_x^2 +V(x)-\mu_s -3\Gamma(x) \phi_s^2(x).\end{array}
\eeq
After multiplication of the first equation in \eqref{E:evp_re} by $\bar{a}$ and of the complex conjugate of the second equation by $b$, integration over $\R$ and addition we obtain 
$$\lambda \|a\|_{L^2}^2+\bar{\lambda}\|b\|_{L^2}^2 = \int_\R-b''(x)\bar{a}(x)+\bar{a}''(x)b(x)+2\Gamma(x)\phi_s^2(x)\bar{a}(x)b(x)dx.$$
Integration by parts leads to 
$$\left|\lambda \|a\|_{L^2}^2+\bar{\lambda}\|b\|_{L^2}^2\right|=2\left|\int_\R \Gamma(x)\phi_s^2(x)\bar{a}(x)b(x)dx\right| \leq 2\|\Gamma\|_\infty\|\phi_s\|_\infty^2\|a\|_{L^2}\|b\|_{L^2},$$ 
where the inequality follows by Cauchy-Schwarz. Finally, for $\text{Re}(\lambda)$ we obtain $|\text{Re}(\lambda)| \leq ||\Gamma||_{\infty} ||\phi_s||_{\infty}^2$ using the inequality $2\alpha \beta\leq \alpha^2+\beta^2$ for $\alpha,\beta \in \R$.
\end{proof}

Earlier works apply a discretization method to equation \eqref{E:linearized} and compute the eigenvalues of the corresponding system directly, see e.g. \cite{SK02}, where GSs were studied. Yet, as pointed out in \cite{PSK04,DG05}, this may be highly computationally costly and typically leads to spurious eigenvalues of \eqref{E:evp}. Identifying those spurious values is a nontrivial and absolutely necessary task when studying stability of solitary waves. The Evans function method was shown in \cite{DG05,PSK04} to avoid this problem. 

\subsection{The Evans Function Method} \label{S:evans_method}

The Evans function $\CE(\lambda), \CE: \C\rightarrow \C$ is a generalization of the characteristic polynomial for a linear operator. Its zeros coincide with isolated eigenvalues of the operator. The multiplicity of the zero equals the algebraic multiplicity of the eigenvalue. The Evans function is defined and analytic away from the essential spectrum of $\L$. It was introduced first by J. Evans in his study of stability of nerve impulses \cite{Evans75}. For analysis of the Evans function see, e.g. \cite{AGJ90,PW92}. It has been used for both analytical \cite{AGJ90,PW92,Sandstede02,KS04} and numerical \cite{DG05,Brin01,AB02,PSK04,GW08} studies of stability of traveling and standing solitary waves. For problems with exponential dichotomy the Evans function is a determinant of a matrix generated by bases of the stable and unstable manifolds corresponding to the trivial solution $u \equiv 0$ of \eqref{E:evp}. The stable and unstable manifolds consist of solutions that decay exponentially fast as $x\rightarrow \infty$ and $x\rightarrow -\infty$, respectively. When this determinant vanishes at a given $\lambda$, the manifolds are linearly dependent, which means that a solution exists which decays exponentially for both $x\rightarrow \infty$ and $x\rightarrow -\infty$, implying $\lambda \in \sigma_{\text{disc}}(\L)$.

In \cite{PSK04} the numerical Evans function method was used in combination with the variation of constants to study linear stability of GSs of \eqref{E:PNLS} (with $\Gamma_+=\Gamma_-$). The authors, however, did not search systematically for eigenvalues over the whole right half plane. Instead they concentrate on real eigenvalues and on eigenvalues bifurcating from the edges of $\sigma_\text{ess}(\L)$. We show, in fact, that for a numerically stable and accurate evaluation of $\CE(\lambda)$ for $|\lambda|$ large a change of variables and the use of exterior algebra are necessary. With the help of these tools, which were not applied in \cite{PSK04}, we carry out winding number computations of $\CE(\lambda)$ with a contour $\gamma = \delta + \ri\R, \ 0<\delta <<1$, parallel to the imaginary axis in order to determine the number of eigenvalues in the whole half plane to the right of $\gamma$. Note that thanks to an asymptotic behavior of $\CE(\lambda)$ for $|\lambda|\rightarrow \infty$, we will be able to use a finite contour.

\subsubsection{Construction of the Evans Function for \eqref{E:evp}}\label{S:Evans_constr}

As advertised above, the Evans function is constructed using the stable and unstable manifolds of the trivial solution $u\equiv 0$ of \eqref{E:evp}. For $|x|\rightarrow \infty$ the terms in \eqref{E:evp} that are proportional to $\phi_s^2(x)$ vanish due to the exponential decay of $\phi_s$ and one obtains the uncoupled Hill's equations
\begin{subequations}\label{E:Hill}
\beq\label{E:Hill_plus}
-\pa_x^2 u_1+V(x)u_1 = \mu_+ u_1
\eeq
\beq\label{E:Hill_minus}
-\pa_x^2 u_2+V(x)u_2 =\mu_- u_2,
\eeq
\end{subequations}
where $\mu_\pm:=\mu_s\pm \ri \lambda$. If $\mu_+\notin \sigma(L)$, the fundamental system of \eqref{E:Hill_plus} consists of the Bloch waves 
\beq\label{E:Bloch_plus}
\psi_{1,2}^{\mu_+}(x)=p_{1,2}^{\mu_+}(x)e^{\pm \ri k_+ x}, \qquad \text{with} \; p_{1,2}^{\mu_+}(x+d)=p_{1,2}^{\mu_+}(x), \quad \text{Im}(k_+)>0,
\eeq
which decay exponentially fast as $x\rightarrow \pm \infty$ respectively. Similarly for $\mu_-\notin \sigma(L)$ the Bloch waves 
\beq\label{E:Bloch_minus}
\psi_{1,2}^{\mu_-}(x)=p_{1,2}^{\mu_-}(x)e^{\pm \ri k_- x} \qquad \text{with} \; p_{1,2}^{\mu_-}(x+d)=p_{1,2}^{\mu_-}(x), \quad  \text{Im}(k_-)>0,
\eeq
that build the fundamental system of \eqref{E:Hill_minus}, decay exponentially fast as $x\rightarrow \pm \infty$ respectively. We choose the following normalization of the Bloch waves in \eqref{E:Bloch_plus},\eqref{E:Bloch_minus}
\beq\label{E:Bloch_normaliz}
p_{1,2}^{\mu_+}(0)=1, \quad  p_{1,2}^{\mu_-}(0)=1.
\eeq
Note that for even potentials $V$ this normalization is possible since $\mu_\pm$ do not lie in $\sigma(L)$ and $p_{1,2}^{\mu_\pm}$ are thus not Dirichlet eigenfunctions of \eqref{E:shifted_evp} (with $k=k_+$ and $k=k_-$ for $p_{1,2}^{\mu_+}$ and $p_{1,2}^{\mu_-}$ respectively) so that $p_{1,2}^{\mu_\pm}(0)\neq 0$. For an even $V$ a Dirichlet eigenfunction $p(x)$ would allow the extension of the solution $\psi(x)=p(x)e^{\ri kx}$ via symmetry from the half line on which $\psi$ is bounded to $\R$, producing so a bounded solution, which is impossible since $\mu_\pm \notin \sigma(L)$.

For $x\rightarrow \infty$ the \textit{stable manifold} of \eqref{E:evp} is thus generated by
\beq\label{E:stab_manif_asymp}
\left\{\bpm\psi_{1}^{\mu_+}\\0\epm, \bpm 0\\ \psi_{1}^{\mu_-}\epm\right\},
\eeq
and for $x\rightarrow -\infty$ the \textit{unstable manifold} of \eqref{E:evp} is generated by
\beq\label{E:unstab_manif_asymp}
\left\{\bpm\psi_{2}^{\mu_+}\\0\epm,\bpm 0\\\psi_{2}^{\mu_-}\epm\right\}.
\eeq
The four vector functions in \eqref{E:stab_manif_asymp} and \eqref{E:unstab_manif_asymp} are asymptotic reductions of the fundamental solution set of \eqref{E:evp} though each of the pairs is for a different asymptotic region. Note that we use the normalization \eqref{E:Bloch_normaliz} also as a normalization of the fundamental solutions of \eqref{E:evp}. 

Let us rewrite \eqref{E:evp} as the first order system 
\beq \label{E:first_order_system}
v'=Av, \qquad A:= \begin{pmatrix} 0 & 1 & 0 & 0 \\ V(x) -2 \Gamma(x) \phi_s^2(x) - \mu_+ & 0 & -\Gamma(x) \phi_s^2(x) & 0 \\ 0 & 0 & 0 & 1 \\ -\Gamma(x) \phi_s^2(x) & 0 & V(x) -2 \Gamma(x) \phi_s^2(x) - \mu_- & 0 \end{pmatrix}
\eeq
for $v:=(u_1,u_1',u_2,u_2')$. The fundamental system of \eqref{E:first_order_system} is denoted by $\{v_1^-,v_2^-,v^+_1,v^+_2\}$. The labels are assigned according to the above asymptotic behavior as follows
\begin{subequations}\label{E:sol_asymp}
\beq\label{E:sol_asymp_min}
v_1^-(x) \sim \begin{pmatrix} \psi_2^{\mu_+}(x) \\ (\psi_2^{\mu_+})'(x) \\ 0 \\ 0 \end{pmatrix}, \quad v_2^-(x) \sim \begin{pmatrix} 0 \\ 0 \\ \psi_2^{\mu_-}(x) \\ (\psi_2^{\mu_-})'(x) \end{pmatrix} \; \text{as} \; x \to -\infty \; , \ \text{and} 
\eeq
\beq\label{E:sol_asymp_plus}
v_1^+(x) \sim \begin{pmatrix} \psi_1^{\mu_+}(x) \\ (\psi_1^{\mu_+})'(x) \\ 0 \\ 0 \end{pmatrix}, \quad  v_2^+(x) \sim \begin{pmatrix} 0 \\ 0 \\ \psi_1^{\mu_-}(x) \\ (\psi_1^{\mu_-})'(x) \end{pmatrix} \; \text{as} \; x \to \infty.
\eeq
\end{subequations}
Note that $\lambda$ is a parameter of $v_{1,2}^\pm$. The Evans function $\CE$ is then defined as
\beq \label{E:Evans}
\CE(\lambda) = \operatorname{det}(v_1^-(x_*; \lambda), v_2^-(x_*; \lambda),v_1^+(x_*; \lambda), v_2^+(x_*; \lambda)).
\eeq
As $\CE$ is constant in $x_*$, its choice is arbitrary and for our purposes of stability of SGSs we choose $x_*=0$, i.e., at the interface location, which avoids the need of numerical integrations of \eqref{E:first_order_system} across the interface. 

The basic (naive) idea of how to evaluate $\CE(\lambda)$ is to solve \eqref{E:first_order_system}  for $v_{1,2}^-$ up to $x=x_*$ with the initial conditions given by \eqref{E:sol_asymp_min} evaluated at $x=-L_\infty<<-1$ and for $v_{1,2}^+$ up to $x=x_*$ with the initial conditions given by \eqref{E:sol_asymp_plus} evaluated at $x=L_\infty$. In Section \ref{S:stab_Evans} we discuss difficulties of this approach.

A-priori information about the behavior of $\CE$ includes the upper bound on the real part of its zeros given by Lemma \ref{L:bound_sp}. Besides, we have the following symmetry of $\CE$ and asymptotic behavior of $\CE(\lambda)$ for $|\lambda|\rightarrow \infty$.
\begin{lemma}\label{L:E_sym}
The Evans function $\CE$ in \eqref{E:Evans} satisfies $\CE(\bar{\lambda})=\overline{\CE(\lambda)}$.
\end{lemma}
\begin{proof}
Complex conjugation of \eqref{E:first_order_system} reveals that $\overline{v(x;\lambda)}$ and $Pv(x;\bar{\lambda})$ satisfy the same differential equation $u'=\overline{A(\lambda)}u$, where $P$ is the permutation matrix $P:=\left(\begin{smallmatrix}0&0&1&0\\0&0&0&1\\1&0&0&0\\0&1&0&0\end{smallmatrix}\right)$. 
Because $\overline{\psi_1^{\mu_+(\lambda)}}(x) = \psi_1^{\mu_-(\bar{\lambda})}(x)$ and $\overline{\psi_2^{\mu_+(\lambda)}}(x) = \psi_2^{\mu_-(\bar{\lambda})}(x)$, the asymptotic data \eqref{E:sol_asymp} are related via complex conjugation and the permutation $P$ in the following way:
$\overline{v_1^-(x,\lambda)} \sim P v_2^-(x,\bar{\lambda})$ as $x\rightarrow -\infty$ and $\overline{v_1^+(x,\lambda)} \sim P v_2^+(x,\bar{\lambda})$ as $x\rightarrow \infty$. We thus have
\[
\begin{split}
\overline{\CE(\lambda)}=&\operatorname{det}(\overline{v_1^-(x_*; \lambda)}, \overline{v_2^-(x_*; \lambda)},\overline{v_1^+(x_*; \lambda)}, \overline{v_2^+(x_*; \lambda)})\\
=&\operatorname{det}\left(P(v_2^-(x_*; \bar{\lambda}), v_1^-(x_*; \bar{\lambda}),v_2^+(x_*; \bar{\lambda}), v_1^+(x_*; \bar{\lambda}))\right)\\
=&\operatorname{det}\left(P(v_1^-(x_*; \bar{\lambda}), v_2^-(x_*; \bar{\lambda}),v_1^+(x_*; \bar{\lambda}), v_2^+(x_*; \bar{\lambda}))P_2\right)=\CE(\bar{\lambda})
\end{split}
\]
where $P_2:=\left(\begin{smallmatrix}0&1&0&0\\1&0&0&0\\0&0&0&1\\0&0&1&0\end{smallmatrix}\right)$
and where the product rule for determinants together with the fact that $\operatorname{det}(P)=\operatorname{det}(P_2)=1$ were used. 
\end{proof}

\begin{lemma}\label{L:E_asympt}
The renormalized Evans function 
\beq\label{E:Evans_renorm}
E(\lambda):=\frac{-1}{4\lambda+1}\CE(\lambda)
\eeq 
is analytic for $\text{Re}(\lambda)>0$ and satisfies $E(\lambda) \rightarrow 1$ as $|\lambda|\rightarrow \infty$ for $\text{arg}(\lambda)\in (-\pi/2, \pi/2)$.
\end{lemma}
\begin{proof}
The analyticity follows from analyticity of ${-1 \over 4\lambda+1}$ and $\CE(\lambda)$. 

For $|\lambda|$ large all terms except those proportional to $\lambda$ become negligible in the second and fourth equations in \eqref{E:first_order_system} leading to 
\beq\label{E:lambda_asymp}
v'=\tilde{A}v \quad \text{with} \quad \tilde{A} = \left(\begin{smallmatrix}0&1&0&0\\-\ri\lambda&0&0&0\\0&0&0&1\\0&0&\ri\lambda&0\end{smallmatrix}\right).
\eeq
The 4 eigenvalues of $\tilde{A}$ are $\pm \sqrt{-\ri\lambda}$ and $\pm \sqrt{\ri\lambda}$ with eigenvectors $(1,\pm\sqrt{-\ri\lambda},0,0)^T$ and $(0,0,1,\pm\sqrt{\ri\lambda})^T$ respectively. The fundamental solution set of \eqref{E:lambda_asymp} can thus be chosen as 
\beq\label{E:fundam_syst_asymp}
\left\{e^{\sqrt{-\ri\lambda}x} \left(\begin{smallmatrix}1\\\sqrt{-\ri\lambda}\\0\\0\end{smallmatrix}\right), e^{\sqrt{\ri\lambda}x} \left(\begin{smallmatrix}0\\0\\1\\ \sqrt{\ri\lambda}\end{smallmatrix}\right),e^{-\sqrt{-\ri\lambda}x} \left(\begin{smallmatrix}1\\-\sqrt{-\ri\lambda}\\0\\0\end{smallmatrix}\right), e^{-\sqrt{\ri\lambda}x} \left(\begin{smallmatrix}0\\0\\1\\ -\sqrt{\ri\lambda}\end{smallmatrix}\right)  \right\}.
\eeq
With the convention that the complex argument of the `+' square root function $\sqrt{z}$ lies in $(-\pi/2,\pi/2]$, i.e., arg$(\sqrt{z})\in(-\pi/2,\pi/2]$,
we get by comparing with the asymptotic behavior in \eqref{E:sol_asymp_min}, \eqref{E:sol_asymp_plus} that the ordering of the vectors \eqref{E:fundam_syst_asymp} agrees with the one in $\CE(\lambda)$. Also the normalization agrees with the one implied by \eqref{E:Bloch_normaliz}. Hence the vectors in \eqref{E:fundam_syst_asymp} can be used directly to evaluate $\CE(\lambda)$ for $|\lambda|$ large.  A calculation of the determinant of the vectors in \eqref{E:fundam_syst_asymp} for $|\lambda|\rightarrow \infty$ yields $\CE(\lambda) \sim -4\lambda$ if $\text{arg}(\lambda)\in (-\pi/2, \pi/2)$ and $\CE(\lambda) \sim 4\lambda$ if $\text{arg}(\lambda)\in (\pi/2, 3\pi/2)$. 
The imaginary axis is excluded to avoid the essential spectrum where our Evans function has not been defined. The renormalized function $E(\lambda)$ thus converges to 1 for $\text{arg}(\lambda)\in (-\pi/2, \pi/2)$.
\end{proof}

\subsection{The Winding Number Method and a Choice of the Contour} \label{S:wind}
The argument principle \cite{Palka91} is a standard tool for counting the number of zeros of a complex function within a region of its analyticity. Denoting by  
$ n(\gamma, z)$ the \textit{winding number} \cite{Palka91} of a closed curve $\gamma\subset \C$ with respect to the point $z\in \C\setminus \gamma$, the \textit{argument principle} states that if $E$ is an analytic function inside $\gamma$ and has no zeros on $\gamma$, then
\beq\label{E:arg_princip}
n(E(\gamma),0)=N,
\eeq
where $N$ is the number of zeros of $E$, counting multiplicities, inside $\gamma$.

Using \eqref{E:arg_princip} we can, in principle, determine the number of eigenvalues of $\L$ in the right half complex plane by using a contour $\gamma$ that encloses it. $\mathcal{E}$ is, however, not analytic on $\sigma_{\text{ess}}(\L) \subset \ri \R$ so that $\gamma$ has to stay a positive distance away from the imaginary axis. Although it should be possible to extend the Evans function analytically across the essential spectrum by generalizing the analysis of \cite{KS98,KS02,KS04} for our periodic coefficient case, the use of the imaginary axis as a contour for the argument principle would still be unpractical due to the expected occurrence of zeros of the extended Evans function within the essential spectrum. This expectation is based on the analysis of constant coefficient problems, for instance, in \cite{KS02}, where zeros in the essential spectrum are found and their bifurcation from spectral edges studied.

We take $\gamma  \parallel \ri \R$ with $\delta:=\text{dist}(\gamma,\ri\R)>0$. In most of the  numerical computations we use $\delta = 0.005$. Due to the $\delta$-gap between $\gamma$ and the imaginary axis, we are thus able to detect only eigenvalues with real part greater than $\delta$, i.e., only `substantial' instabilities. Using the symmetry from Lemma \ref{L:E_sym} and the asymptotic result from Lemma \ref{L:E_asympt}, the curve $\gamma$ can in practice be replaced by
\beq\label{E:contour}
\tilde{\gamma}=\delta +\ri[0,H],
\eeq
where $H>0$ is chosen for a given $\phi_s$ so that near $\text{Im}(\tilde{\gamma})=H$ the Evans function $E(\lambda)$ is clearly converging to $1$. In most of our computations $H=20$ is sufficient. In order to numerically evaluate the winding number, we have written a simple Matlab script, see Appendix \ref{A:wind_code}.
  
It is possible that $E$ has a zero on $\tilde{\gamma}$, which would prevent the applicability of the argument principle. In numerics this will, however, generically result in a small value of $E(\lambda)$ so that the effect is the same as a zero near $\tilde{\gamma}$. If $E$ attains numerically the value $0$ somewhere along $\tilde{\gamma}$, the contour can be locally deformed. This case does, however, not occur in our simulations. 

Reliable computations of $n(E(\gamma),0)$ require a sufficiently fine discretization of $\tilde{\gamma}$. Needlessly fine discretization, on the other hand, lead to intolerably long computation times. A possible solution is the adaptive discretization suggested in Sec. 3.3 of \cite{Brin01} with the main idea being to ensure that the difference in the complex argument $\text{arg}(E(\lambda))$ for neighboring points $\lambda_{1,2}\in \tilde{\gamma}$ is near a desired value. 



\subsection{Numerically Stable Evaluation of $E(\lambda)$}\label{S:stab_Evans}

Attempting a straightforward evaluation of $E(\lambda)$ as defined via \eqref{E:first_order_system} - \eqref{E:Evans_renorm}, one encounters several issues with numerical stability and accuracy. Firstly, the evaluation of the initial conditions \eqref{E:sol_asymp} at some $x=\pm L_\infty, L_\infty >>1$ requires computation of exponentially growing/decaying Bloch functions, where the growth/decay rate is very strong for certain parameter values. Secondly,  
the solutions of \eqref{E:first_order_system} also feature exponential growth/decay. In both of these cases parameter dependent transformations can be used to remove the exponential growth/decay. Next, more importantly, the system \eqref{E:first_order_system} suffers from stiffness when the growth/decay rates of the two vectors in the manifold of interest (stable or unstable) are largely different. This can be overcome by a reformulation of \eqref{E:first_order_system} in the exterior algebra. As explained below, the use of Grassmanian preserving ODE integrators is then vital. In the following we describe these problems and their solutions in detail.

\subsubsection{Computation of the Bloch Functions in \eqref{E:sol_asymp}}\label{S:Bloch_fn_comput}

Due to the form \eqref{E:Bloch_plus}, \eqref{E:Bloch_minus} the Bloch functions $\psi_{1,2}^{\mu_\pm}$ need to be, of course, computed only on the period $x\in [0,d]$. In  \eqref{E:sol_asymp} their values at $x=\pm L_\infty, L_\infty>>1$, which are used as initial data for \eqref{E:first_order_system}, are then obtained straightforwardly due to periodicity of $p_{1,2}^{\mu_\pm}$. Two commonly used methods are available to compute the Bloch functions. The first possibility is to solve \eqref{E:shifted_evp} with $\omega(k)=\mu_\pm$ fixed as a quadratic eigenvalue problem in $(k,p^{\mu_\pm})$ on $x\in[0,d]$ with periodic boundary conditions. To avoid the need for solving quadratic eigenvalue problems, we opt for the second traditional method, which is Floquet theory \cite{MagWin66,Eastham73}, in which \eqref{E:Hill_plus} and \eqref{E:Hill_minus} are solved on $[0,d]$ as initial value problems and $k_\pm$ are determined from eigenvalues of the monodromy matrix, i.e., from Floquet multipliers. Even on the bounded domain $[0,d]$ the exponential growth of $\psi^{\mu_\pm}_{1,2}$, however, leads to a strong enhancement of error when $|\text{Im}(k_\pm)|$ is large. As shown below, this occurs, in particular, for $|\mu_\pm|$ large when $\text{arg}(\mu_\pm)$ is not close to $\pi$ or $-\pi$.

For $|\mu|$ large the equation $-\pa_x^2 w +V(x) w = \mu w$ reduces effectively to 
\[-\pa_x^2 w =\mu w\]
with the fundamental system $\{e^{\pm \sqrt{-\mu}x}\}.$ With the definition of the `+' square root $\sqrt{z}, z\in \C$ such that $\text{arg}(\sqrt{z}) \in (-\pi/2,\pi/2]$, the solution $e^{\sqrt{-\mu}x}$ grows fast when $|\mu|$ is large and $\text{arg}(\mu)$ is not close to $\pm \pi$. In \eqref{E:Hill} and with the choice of the contour along the upper half of the imaginary axis as in \eqref{E:contour} this happens for $\mu_- \ (=\mu_s-\ri \lambda)$ when $|\lambda|$ is large. In \eqref{E:Bloch_minus} we thus get a large $\text{Im}(k_-)$. For $\mu_+ \ (=\mu_s+\ri \lambda)$ the contour \eqref{E:contour} leads to $\text{arg}(\mu_+)$ near $\pi$ for $|\mu_+|$ large so that no strong growth occurs in $\psi_{1,2}^{\mu_+}$, i.e., $\text{Im}(k_-)$ is close to $0$.

To achieve sufficient accuracy of $\psi^{\mu_-}_{1,2}$, we use a change of variables that removes the exponential growth. Rewriting \eqref{E:Hill_minus} as a first order system for $v:=(u_2,u_2')^T$, we have
\begin{equation}\label{E:Hill_1st_ord}
v'=Bv, \qquad B := \begin{pmatrix} 0 & 1 \\ V(x)-\mu_- & 0 \end{pmatrix}.
\end{equation}
The Bloch functions of \eqref{E:Hill_1st_ord} are
\begin{equation*}
v^{(1)}(x) = \left( \psi^{\mu_-}_1(x), (\psi^{\mu_-}_1)'(x) \right)^T, \quad v^{(2)}(x) = \left(\psi^{\mu_-}_2(x), (\psi^{\mu_-}_2)'(x) \right)^T,
\end{equation*}
which due to \eqref{E:Bloch_minus} have the form
\begin{equation}\label{E:v1_v2}
v^{(1)}(x) = q^{(1)}(x) e^{\ri k_- x}, \ 
v^{(2)}(x) = q^{(2)}(x) e^{-\ri k_- x}, 
\end{equation}
where
\begin{equation}\label{E:q1_q2}
q^{(1)}(x) = \begin{pmatrix} p^{\mu_-}_1(x) \\ (p^{\mu_-}_1)'(x) + \ri k_- p^{\mu_-}_1(x) \end{pmatrix} \; \ \text{and} \quad \; q^{(2)}(x) = \begin{pmatrix} p^{\mu_-}_2(x) \\ (p^{\mu_-}_2)'(x) - \ri k_- p^{\mu_-}_2(x) \end{pmatrix}.
\end{equation}
To remove the exponential growth, we introduce $\tilde{v}^{(j)}(x):=v^{(j)}(x)e^{-\sqrt{-\mu_-} x}$ such that
\begin{align} \label{E:v_til_equ}
\tilde{v}' = (B - \sqrt{-\mu_-} I) \tilde{v}.
\end{align}
Via standard Floquet theory \cite{Eastham73} the Bloch functions 
\beq\label{E:Bloch_fns_til}
\tilde{v}^{(1,2)}(x) = q^{(1,2)}(x)e^{\ri \tilde{k}_{1,2}x}
\eeq
of \eqref{E:v_til_equ} can be computed by solving \eqref{E:v_til_equ} on $x\in[0,d]$ and finding the eigenvalues of the monodromy matrix, i.e. the characteristic multipliers $\tilde{\rho}_{1,2}=e^{\ri \tilde{k}_{1,2}d}$. As a `side effect' the change of variables $v \rightarrow \tilde{v}$ has doubled the decay rate in $v^{(1)}$, which possibly results in extremely small values of $\tilde{v}^{(1)}$ near $x=d$ such that for $|\mu_-|$ large it is numerically impossible to recover accurate information about $\tilde{v}^{(1)}$. We, therefore, use \eqref{E:v_til_equ} to compute only $v^{(2)}$. The corresponding $\tilde{v}^{(2)}$ is identified as the solution belonging to the larger characteristic multiplier, i.e., $|\tilde{\rho}^{(2)}|>|\tilde{\rho}^{(1)}|$, and $\tilde{k}_2$ is then obtained from $\tilde{k}_2 = -\ri \log(\tilde{\rho}^{(2)})/d$. The periodic part $q^{(2)}$ is computed from \eqref{E:Bloch_fns_til}. Finally, $k_-$ is given by
\[k_- = \ri \sqrt{-\mu_-}-\tilde{k}_2.\]

The missing Bloch function $v^{(1)}$, or equivalently $\psi_1^{\mu_-}$, is readily available via symmetry. For $\psi_{1,2}^{\mu_-} = p_{1,2}^{\mu_-}(x)e^{\pm\ri k_- x}$ equation \eqref{E:Hill_minus} becomes
\begin{subequations}\label{E:per_Bloch_prob}
\beq\label{E:per_Bloch_1}
-(\partial_x + \ri k_-)^2 p^{\mu_-}_1 + V(x) p^{\mu_-}_1 = \mu_- p^{\mu_-}_1, 
\eeq
\beq\label{E:per_Bloch_2}
-(\partial_x - \ri k_-)^2 p^{\mu_-}_2 + V(x) p^{\mu_-}_2 = \mu_- p^{\mu_-}_2 
\eeq
\end{subequations}
with periodic boundary conditions on $x\in[0,d]$.
The substitution $k_- \rightarrow -k_-$ and $x\rightarrow -x$ in the second equation reveals that 
$$p_2^{\mu_-}(x)=p_1^{\mu_-}(-x).$$
Note that $p_1^{\mu_-}$ is unique up to a normalization, i.e., $\mu_-$ is a simple eigenvalue of \eqref{E:per_Bloch_1} since otherwise for $k_-\neq 0$ one could generate via the above symmetry at least four linearly independent solutions of  \eqref{E:Hill_minus}. The case $k_-=0$ yields a periodic Bloch function and can thus only occur for $\mu_-\in \sigma(L)$ \cite{MagWin66,Eastham73}, which is not our case since $\text{Im}(\mu_-)=-\text{Re}(\lambda)\neq 0$

\subsubsection{Dealing with the Stiffness of \eqref{E:first_order_system}}

In Sec. \eqref{S:Evans_constr} the Evans function is defined using the bases $\{v_1^-,v_2^-\}$ and $\{v_1^+,v_2^+\}$ of the unstable and stable manifolds respectively of the trivial solution of  \eqref{E:first_order_system}. In practice $v^-_{1,2}$ need to be computed from $x=-L_\infty << -1$ up to $x=x_*$ and $v^+_{1,2}$ from $x=L_\infty >> 1$ down to $x=x_*$.  In the unstable manifold $v_1^-$ and $v_2^-$ have different growth rates. For large $|x|$, where the terms proportional to $\phi_s^2(x)$ are negligible, the rates are $e^{\text{Im}(k_+)x}$ and $e^{\text{Im}(k_-)x}$ respectively. When $\text{Im}(k_+)$ and $\text{Im}(k_-)$ differ strongly, the problem is stiff and the fast growing solution dominates the dynamics and in a numerical simulation the error is enhanced in the direction of this more unstable mode so that the solution growing more slowly cannot be accurately computed. The functions $v_1^-$ and $v_2^-$ thus do not remain linearly independent in the numerical $x-$evolution although they are linearly independent initially, see \eqref{E:sol_asymp_min}. An analogous situation occurs for $v_1^+$ and $v_2^+$  of the stable manifold. We saw in Sec. \ref{S:Bloch_fn_comput} that for the selected vertical contour this large difference of growth rates occurs, in particular, for $\text{Im}(\lambda)$ large, where $\text{Im}(k_-)$ is large and $\text{Im}(k_+)$ is close to zero.

A traditional approach to a problem with linear dependence is to apply orthogonalization. Orthogonality for all $x$ is, however, not generally satisfied by solutions of 
\eqref{E:first_order_system}, \eqref{E:sol_asymp}, and as explained in \cite{AB02,DG05}, orthogonalization may destroy the analyticity of $E=E(\lambda)$. This is inadmissible when we intend to employ the argument principle. An alternative is the use of exterior algebra as first proposed in \cite{NR79} for solving stiff linear ODE systems. More recently it has been used in spectral problems, for example, in \cite{Brin01,AB02,DG05,GW08}. For theoretical background on exterior algebra see, e.g., \cite{GH78,CCL99,Yok92}.

Let $\{e_1, \ldots, e_n\}$ be an orthonormal basis for $\C^n$. We first define the \textit{wedge} (exterior) \textit{product} via the properties of distributivity and associativity
\[
\begin{split}
e_i \wedge (ae_j+e_k) =&  ae_i \wedge e_j + e_i\wedge e_k\\
(ae_i +e_j)\wedge e_k =& ae_i\wedge e_j + e_j\wedge e_k 
\end{split}
\]
and antisymmetry
\[e_i \wedge e_j = - e_j \wedge e_i.\]
Then the distinct and nonzero members of $\{ e_{i_1} \wedge \cdots \wedge e_{i_k}: 1 \leq i_1 < \cdots < i_k \leq n, 1 \leq k \leq n\}$ form a basis of the vector space of all $k$-forms $a^{(1)}\wedge \ldots \wedge a^{(k)}$ with $a^{(j)}\in \C^n, j = 1,\ldots, k$, which is generally denoted by $\bigwedge^{k}(\C^n)$ and has dimension $\binom{n}{k}$. Let $b^{(j)} = \sum_{i=1}^n b_{ij} e_i$ with $j=1,\cdots,n$, and $B := (b_{ij})$. Then (see \cite{Yok92}, Proposition 3.3) 
\beq\label{E:det_rel}
b^{(1)} \wedge \ldots \wedge b^{(n)} = \operatorname{det}B \; e_1 \wedge \ldots \wedge e_n.
\eeq

For example, for $a,b\in\C^n, \ a=\sum_{i=1}^{n}a_ie_i, b = \sum_{i=1}^{n}b_ie_i$ one thus has
\[a\wedge b = \sum_{i=1}^{n-1}\sum_{j=i+1}^n(a_ib_j-a_jb_i)e_i\wedge e_j.\]

Furthermore, for $u\in \bigwedge^j(\C^n)$ and $v\in \bigwedge^k(\C^n)$ one obtains $u\wedge v \in \bigwedge^{j+k}(\C^n)$. In particular, for $j=n-k$ we get $u\wedge v \in \bigwedge^{n}(\C^n)$. In this case (see \cite{AB02}, Section 5)
\beq\label{E:wedge_complem}
u \wedge v =\langle \bar{u},\Sigma v\rangle e_1 \wedge \ldots \wedge e_n,
\eeq
where $\langle\cdot, \cdot\rangle$ is the standard inner product on $\C^{\binom{n}{k}}$ with conjugation of the first argument and $\Sigma$ is an orthogonal matrix. 

Applying \eqref{E:det_rel} to our Evans function problem with $k=2, n=4$, we get
\[
\begin{split}
E(\lambda) e_1 \wedge \ldots \wedge e_4 &= {-1\over 4\lambda+1} v_1^-(x_*;\lambda)\wedge v_2^-(x_*;\lambda)\wedge v_1^+(x_*;\lambda)\wedge v_2^+(x_*;\lambda)\\
&={-1\over 4\lambda+1} V^-(x_*;\lambda)\wedge V^+(x_*;\lambda),
\end{split}
\]
where $V^\pm :=v_1^\pm \wedge v_2^\pm$. Because $V^\pm \in \bigwedge^2(\C^4)$, the rule \eqref{E:wedge_complem} applies and
\beq\label{E:Evans_wedge}
E(\lambda) = {-1\over 4\lambda+1} \langle \overline{V^-},\Sigma V^+\rangle.
\eeq
Indeed, $E(\lambda) = 0$ if and only if $v_1^-, v_2^-, v_1^+,$ and $v_2^+$ are linearly dependent (see e.g. \cite{Yok92}, Lemma 3.4). In \cite{AB02} it is shown that the form of $\Sigma$ for this case is 
\beq\label{E:Sigma}
\Sigma = \left(\begin{smallmatrix}
0&0&0&0&0&1\\
0&0&0&0&-1&0\\
0&0&0&1&0&0\\
0&0&1&0&0&0\\
0&-1&0&0&0&0\\
1&0&0&0&0&0
\end{smallmatrix}\right)
\eeq
provided the basis for $ \bigwedge^2(\C^4)$ has been selected as $e_1 \wedge e_2, e_1\wedge e_3, e_1\wedge e_4, e_2\wedge e_3, e_2\wedge e_4, e_3\wedge e_4$.

Evolution equations for $V^\pm$ are obtained by linearity of the wedge product
\[V^{\pm'} =v_1^{\pm'}\wedge v_2^\pm + v_1^\pm\wedge v_2^{\pm'}=Av_1^\pm\wedge v_2^\pm+v_1^\pm\wedge Av_2^\pm.\]
As described in Section 2 of \cite{AB02}, denoting by $a_{jk}$ the $(j,k)$ entry of $A$, this induces the matrix
\[A^{(2)} = \left(\begin{smallmatrix}
a_{11}+a_{22}&a_{23}&a_{24}&-a_{13}&-a_{14}&0\\
a_{32}&a_{11}+a_{33}&a_{34}&a_{12}&0&-a_{14}\\
a_{42}&a_{43}&a_{11}+a_{44}&0&a_{12}&a_{13}\\
-a_{31}&a_{21}&0&a_{22}+a_{33}&a_{34}&-a_{24}\\
-a_{41}&0&a_{21}&a_{43}&a_{22}+a_{44}&a_{23}\\
0&-a_{41}&a_{31}&-a_{42}&a_{32}&a_{33}+a_{44}\\
\end{smallmatrix}\right),\]
such that
\beq\label{E:wedge_ODE}
V^{\pm'}=A^{(2)}V^\pm.
\eeq

An alternative to solving \eqref{E:first_order_system} for each of $v_1^-$ and $v_2^-$ from $x=-L_\infty$ to $x=x_*$ and $v_1^+$ and $v_2^+$ from  $x=L_\infty$ to $x=x_*$  is thus to solve \eqref{E:wedge_ODE} for $V^-$ from $x=-L_\infty$ to $x=x_*$ and for $V^+$ from  $x=L_\infty$ to $x=x_*$. The 2-dimensional manifolds of \eqref{E:first_order_system} are represented by lines in \eqref{E:wedge_ODE} and in this way we avoid the linear dependence problem discussed above. The initial conditions for \eqref{E:wedge_ODE} at $x=\pm L_\infty$ are determined by the wedge product of respective vectors in \eqref{E:sol_asymp}.

\paragraph{Decomposability and preservation of the Grassmanian.}

It is not true that every $k-$form can be written as a single wedge product of $k$ elements of $\C^n$. In other words not every $k-$form represents a $k-$dimensional subspace of $\C^n$. If this is satisfied, i.e., if $U\in \bigwedge^k(\C^n)$ can be written as
\[U = u^{(1)}\wedge \ldots \wedge u^{(k)}\]
with $u^{(j)}\in \C^n, j = 1, \ldots, k$, then $U$ is \textit{decomposable} \cite{GH78,CCL99,AB02}. 

In our setting $V^\pm$ are decomposable initially at $x=\pm L_\infty$ by construction, but in a numerical integration of \eqref{E:wedge_ODE} this property may not be preserved so that at $x=x_*$ the computed solution no longer represents the 2-dimensional stable or unstable manifold respectively. Restricting to $n=4,k=2$, the following result holds. A 2-form $U\in \bigwedge^2(\C^4)$ is decomposable if and only if $U\wedge U =0$ \cite{GH78}. For the coefficients it means
\[I(U):={1\over 2} \langle \bar{U}, \Sigma U \rangle= U_1U_6-U_2U_5+U_3U_4=0.\]
The quadratic quantity $I(U)$ is called the \textit{Grassmanian}. It is a strong invariant of \eqref{E:wedge_ODE} if $\Sigma A^{(2)}+(\Sigma A^{(2)})^T=0$, see \cite{AB02}, which holds in our case, where $A^{(2)}$ is constructed from $A$ defined in \eqref{E:first_order_system}.

\renewcommand\arraystretch{1.2}

A number of methods are available that preserve strong quadratic invariants \cite{HLW06}.
As discussed in Section \ref{S:remove_exp}, we, however, solve \eqref{E:wedge_ODE} after the change of variables 
\beq\label{E:var_change}
\tilde{V}^\pm:=e^{\pm \nu x}V^\pm
\eeq
with $\nu = \text{Im}(k_+)+\text{Im}(k_-)$. With \eqref{E:var_change} we have $I(\tilde{V}^\pm) = I(V^\pm)e^{\pm 2\nu x}$ and $\tfrac{d}{dx}(I(\tilde{V}^\pm)) = \tfrac{d}{dx}(I(V^\pm))\pm 2\nu I(\tilde{V}^\pm) = \pm 2\nu I(\tilde{V}^\pm)$ so that $I$ is only a weak invariant of the equation for $\tilde{V}^\pm$, namely equation \eqref{E:Vtil_syst}. This means that if $I(\tilde{V}^\pm(x_0))=0$ at some initial point $x_0$, then $I(\tilde{V}^\pm(x)) = 0 \ \forall x\in \R$. Weak invariants are generally impossible to numerically preserve exactly. Approximate preservation can be achieved by projection methods \cite{HLW06}. In our case the numerical solution is projected onto the manifold $\CM:=\{U\in \C^6 : I(U)=0\}$. We solve  \eqref{E:Vtil_syst} via the standard 4-5th order Runge-Kutta method of Matlab (ode45) and execute the projection at regular $x-$intervals, usually of length $\delta_P=0.2$ (i.e. the frequency of the projections within one period of the potential $V$ is $d/\delta_P=50$), see Sec. \ref{S:evans_accur} for more details on the choice of $\delta_P$.

Renaming the solution vector of \eqref{E:Vtil_syst} to $w$, the projection constitutes the following optimization problem. Given a numerical approximation $\tilde{w}$ (at some point $x$), find $w\in \CM$ such that
\[\|w-\tilde{w}\| = \min_{U\in \CM}\|U-\tilde{w}\|,\]
where $\|\cdot\|$ denotes the Euclidean norm in $\C^6$. Defining the Lagrange function $\|w-\tilde{w}\|^2/2-\zeta I(w)$, we get the following nonlinear problem for the Lagrange multiplier $\zeta$
\[I(\tilde{w}+\zeta I'(\tilde{w})^T)=0,\]
which we solve via Newton's iteration
\[\zeta^{(n+1)} = \zeta^{(n)}-\left[\tfrac{d}{d\zeta}(I(\tilde{w}+\zeta I'(\tilde{w})^T))(\zeta^{(n)})\right]^{-1}I(\tilde{w}+\zeta^{(n)} I'(\tilde{w})^T),\]
where 
\[\tfrac{d}{d\zeta}(I(\tilde{w}+\zeta I'(\tilde{w})^T))(\zeta) = \tilde{w}_1^2+2\zeta\tilde{w}_1\tilde{w}_6+\tilde{w}_6^2+\tilde{w}_2^2-2\zeta\tilde{w}_2\tilde{w}_5+\tilde{w}_5^2+\tilde{w}_3^2+2\zeta\tilde{w}_3\tilde{w}_4+\tilde{w}_4^2.\]

In Section \ref{S:evans_accur} we show that the accuracy of the Evans function increases with increasing projection frequency $d/\delta_P$.


\subsubsection{Removing the Exponential Growth in \eqref{E:wedge_ODE}}\label{S:remove_exp}
 
Based on the asymptotic behavior \eqref{E:sol_asymp} of the solutions $v_{1,2}^\pm$ and linearity of the wedge product, we get that $V^-$ behaves like $e^{(\text{Im}(k_+)+\text{Im}(k_-))x}$ for $x$ large negative and $V^+$ behaves like $e^{-(\text{Im}(k_+)+\text{Im}(k_-))x}$ for $x$ large positive. This exponential growth in the backward and forward $x$ direction for $V^\pm$ respectively leads to difficulties with numerical accuracy when $\text{Im}(k_+)+\text{Im}(k_-)$ is large. Clearly, for values of $\lambda$ close to zeros of $E$ the growth is eliminated in the respective opposite asymptotic regions (at $x>>1$ for $V^-$ and at $x<<-1$ for $V^+$) as the stable and unstable manifolds are close to being linearly dependent. In the intermediate part of the domain (e.g. near $x=x_*$) the growth is possibly largely limited. At other values of $\lambda$ the growth may, however, remain large for all $x$. We, therefore, perform the numerics of \eqref{E:wedge_ODE} under the transformation
 \[\tilde{V}^\pm:=e^{\pm(\text{Im}(k_+)+\text{Im}(k_-))x}V^\pm.\]
 The new variables satisfy
 \beq\label{E:Vtil_syst}
 \tilde{V}^{\pm'}=\left(A^{(2)} \pm (\text{Im}(k_+)+\text{Im}(k_-))I\right)\tilde{V}^\pm.
 \eeq
Linearity of the wedge product yields
\[E(\lambda) = {-1\over 4\lambda+1} V^-(x_*;\lambda)\wedge V^+(x_*;\lambda) =  {-1\over 4\lambda+1} \tilde{V}^-(x_*;\lambda)\wedge \tilde{V}^+(x_*;\lambda).\]

\subsection{Numerical Results on Stability of SGSs}\label{S:stab_numerics}

This section presents results of Evans function computations for the spectral problem corresponding to linear stability of the SGS families in Figs. \ref{F:SGS_on} and \ref{F:SGS_off}. In several examples the NLS \eqref{E:PNLS} is then also evolved numerically in time with the selected SGS profile as the initial condition in order to demonstrate the stable or unstable behavior. 

\subsubsection{Accuracy of the Numerical Evans Function}\label{S:evans_accur}
The problem at hand offers a natural test, namely $E(0)=0$, which holds because $0$ is an eigenvalue of $\L$ and $0\notin \sigma_\text{ess}(\L)$. This kernel is caused by the phase invariance of \eqref{E:PNLS}. In the numerics we use $|E(0)|$ as a measure of accuracy of the numerical Evans function. There is a number of factors that may affect accuracy: (i) error in computing the Bloch functions of \eqref{E:v_til_equ}; (ii) numerical error in solving \eqref{E:wedge_ODE} for the two manifolds; (iii) size $2L_\infty$ of the computational domain for \eqref{E:wedge_ODE};  (iv) norm of the residual in computing the SGS profile $\phi_s$; (v) $h$ and $p$ in the FEM computation of the SGS profile; and (vi) frequency $d/\delta_P$ of projections onto the Grassmanian manifold. Regarding (i) and (ii), we use Matlab's built-in Runge-Kutta function ODE45 with adaptive step size control and set the absolute and relative tolerance to $10^{-6}$ and $10^{-8}$ respectively. For (iii) the value of $L_\infty$, at which the initial conditions for $\tilde{V}^\pm$  in \eqref{E:Vtil_syst} are specified, is selected so that $|\phi(x)|<10^{-14}$ for $|x|> L_\infty$. The values for our 16 families are 
\begin{tabular}{lllll|llllll}
$A_{\text{on}}$: & $L_\infty=176$,& \quad & $E_{\text{on}}$: & $L_\infty=155$, \quad& \qquad & $A_{\text{off}}$: & $L_\infty=138$,& \quad & $E_{\text{off}}$: & $L_\infty=135$,\\
$B_{\text{on}}$: & $L_\infty=75$,& \quad & $F_{\text{on}}$: & $L_\infty=270$,\quad& \qquad & $B_{\text{off}}$: & $L_\infty=103$,& \quad & $F_{\text{off}}$: & $L_\infty=260$,\\
$C_{\text{on}}$: & $L_\infty=200$,& \quad & $G_{\text{on}}$: & $L_\infty=110$,\quad& \qquad & $C_{\text{off}}$: & $L_\infty=140$,& \quad & $G_{\text{off}}$: & $L_\infty=118$,\\
$D_{\text{on}}$: & $L_\infty=110$,& \quad & $H_{\text{on}}$: & $L_\infty=324$,\quad& \qquad & $D_{\text{off}}$: & $L_\infty=125$,& \quad & $H_{\text{off}}$: & $L_\infty=250$. \\
\end{tabular}

\begin{figure}
\begin{center}
\epsfig{figure = 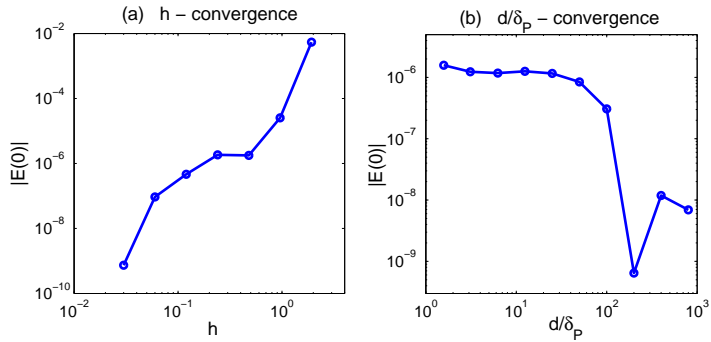 ,scale=0.5}
\end{center}
\caption{Convergence of $|E(0)|$ in dependence on the FEM discretization parameter $h$ in the discretization of \eqref{E:SPNLS} and on the frequency $d/\delta_P$ of projections of the solution of \eqref{E:wedge_ODE} onto the Grassmanian manifold. The selected SGS is one at $\Gamma_-\approx 0.48$ in the family $E_\text{off}$. In (a) we used $d/\delta_P = 200$ and in (b) $h=0.03$.}
\label{F:Eof0_conv}
\end{figure}

For (iv) we use the tolerance $10^{-10}$ on the $L^2$ norm of the residual of \eqref{E:SPNLS}. According to our tests $|E(0)|$ is insensitive to these parameters of (i-iv) if further refined (tolerances decreased and $L_\infty$ increased). We study next sensitivity to (v) and (vi). In both cases we observe a clear convergence of $|E(0|$ with refinement. Fig \ref{F:Eof0_conv} shows dependence of $|E(0)|$ on $h$ in the discretization of \eqref{E:SPNLS} and on the projection frequency $d/\delta_P$ for the case of an SGS at $\Gamma_-\approx 0.48$ in the family $E_\text{off}$ with $p=3$ fixed.


In our computations we use $p=3, h=0.06, d/\delta_P = 50$ as the default for the discretization of \eqref{E:SPNLS}. We reduce $h$ and/or increase $d/\delta_P$ until $|E(0)|<10^{-8}$. This threshold test cannot always be passed in the vicinity of folds where only $10^{-6}$ is often achieved. In such cases it is checked that $|E(0)|< \tfrac{1}{10}\min_{\lambda \in \gamma} |E(\lambda)|$ for the selected contour $\gamma$ as an a-posteriori test. This criterion is satisfied in all the computations except when a zero of $E$ lies in the vicinity of $\gamma$ for the particular SGS.

Note also that with a finite difference (FD) discretization of \eqref{E:SPNLS} the values of $|E(0)|$ are much larger and neither convergence in $dx$ nor in $d/\delta_P$ has been typically observed, which has been tested with both 4th and 2nd order centered FD stencils. FD discretization is exptected to fail due to the profile $\phi$ being only $C^1$ at the interface $x=0$.


\subsubsection{Winding Number Computations}\label{S:wind_nr_results}

We present first winding number computations for the spectral problem as discussed in Sec. \ref{S:wind}. In this section we do not attempt to determine the location of individual eigenvalues (if any) in the right half complex plane.

The selection of the contour and its discretization is described in Sec. \ref{S:wind}. We stress again that due to the small gap between the contour $\gamma$ and the imaginary axis we are able to capture only eigenvalues with $\text{Re}(\lambda)> \text{Re}(\gamma)$. In  all the computations in Fig. \ref{F:SGS_on_stab} and \ref{F:SGS_off_stab} $\text{Re}(\gamma)=0.005$ was used. In Fig. \ref{F:SGS_stab_near_im} several examples with a smaller distance from the imaginary axis are also presented.

Equation \eqref{E:wedge_ODE} was solved up to $x=0$ for both $V^+$ and $V^-$, i.e., up to the interface location. The Evans function was thus evaluated at $x_\star = 0$. After the evaluation of $E$ along $\gamma$, the winding number $n(E(\gamma),0)$ was computed numerically using a simple Matlab script, see Appendix \ref{A:wind_code}.

\begin{figure}
\begin{center}
\epsfig{figure = 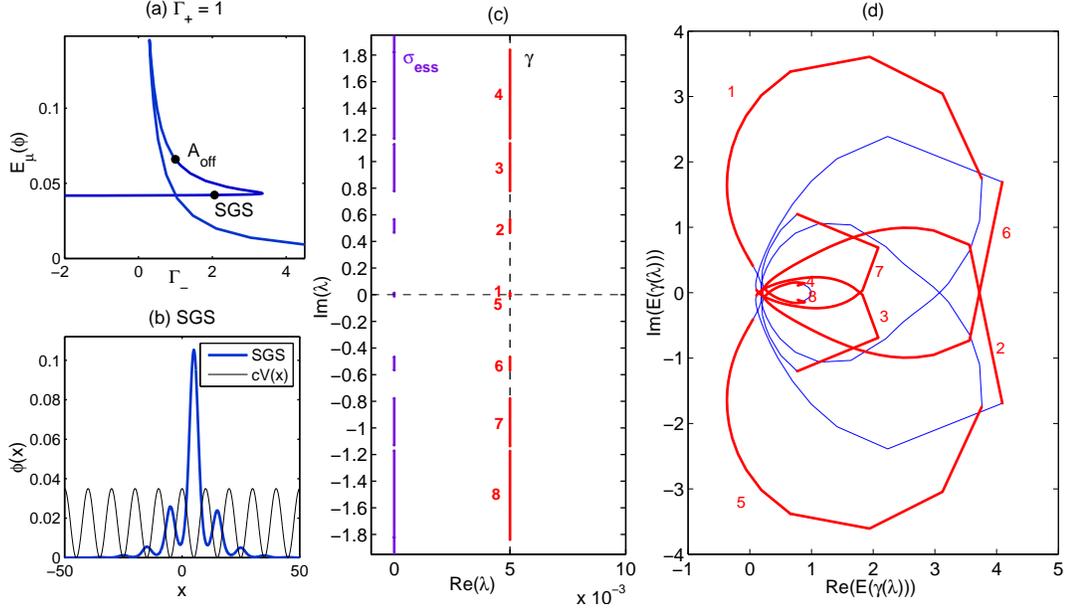,scale=.5}
\end{center}
\caption{Evans function graph for a selected SGS. (a) The SGS family $A_\text{off}$, a selected SGS at $\Gamma_- \approx 2.063$ marked; (b) Profile of the SGS marked in (a); (c) the essential spectrum $\sigma_{\text{ess}}(\L)$ (lie segments along $\text{Re}(\lambda)=0$) and the contour $\gamma$ (vertical dashed line) with segments close to $\sigma_{\text{ess}}(\L)$ marked in bold red; (d) $E(\gamma)$ for the SGS plotted in (b). Bold red segments mark the image of the marked segments along $\gamma$ in (c).}
\label{F:Evans_fn_plot}
\end{figure}

\begin{figure}
\begin{center}
\epsfig{figure = 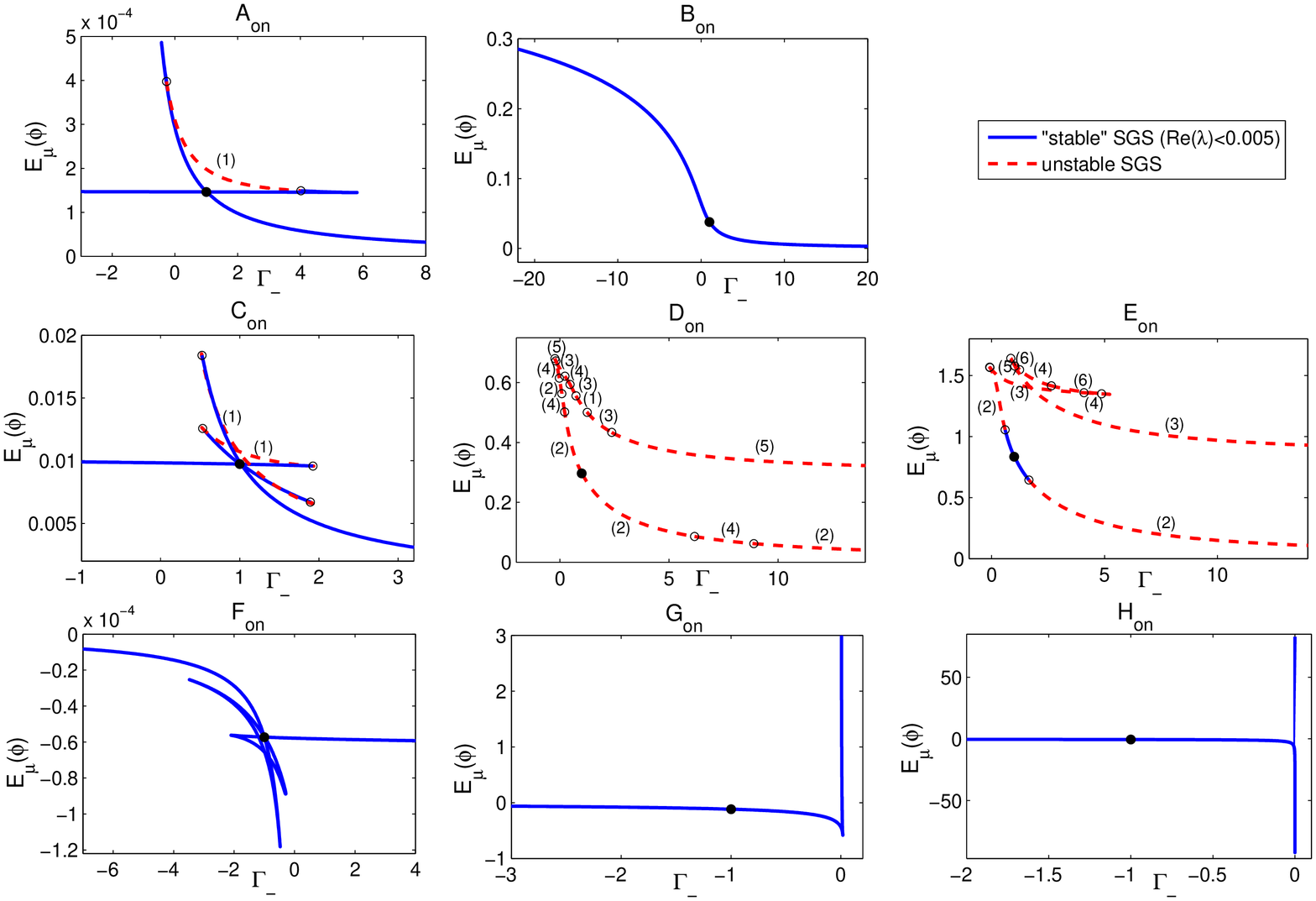,scale=.5}
\end{center}
\caption{Evans function winding number computations for selected SGS families bifurcating from onsite GSs. The vertical contour $\gamma$ with $\text{Re}(\gamma)=0.005$ was used. Full (blue) lines mark `stable' SGSs, where $n(E(\gamma),0)=0$ and dashed (red) lines unstable SGSs, where $n(E(\gamma),0)\neq 0$. Changes in the values of $n$ are marked by black circles and the value is written next to each segment of the curve.}
\label{F:SGS_on_stab}
\end{figure}

\begin{figure}
\begin{center}
\epsfig{figure = 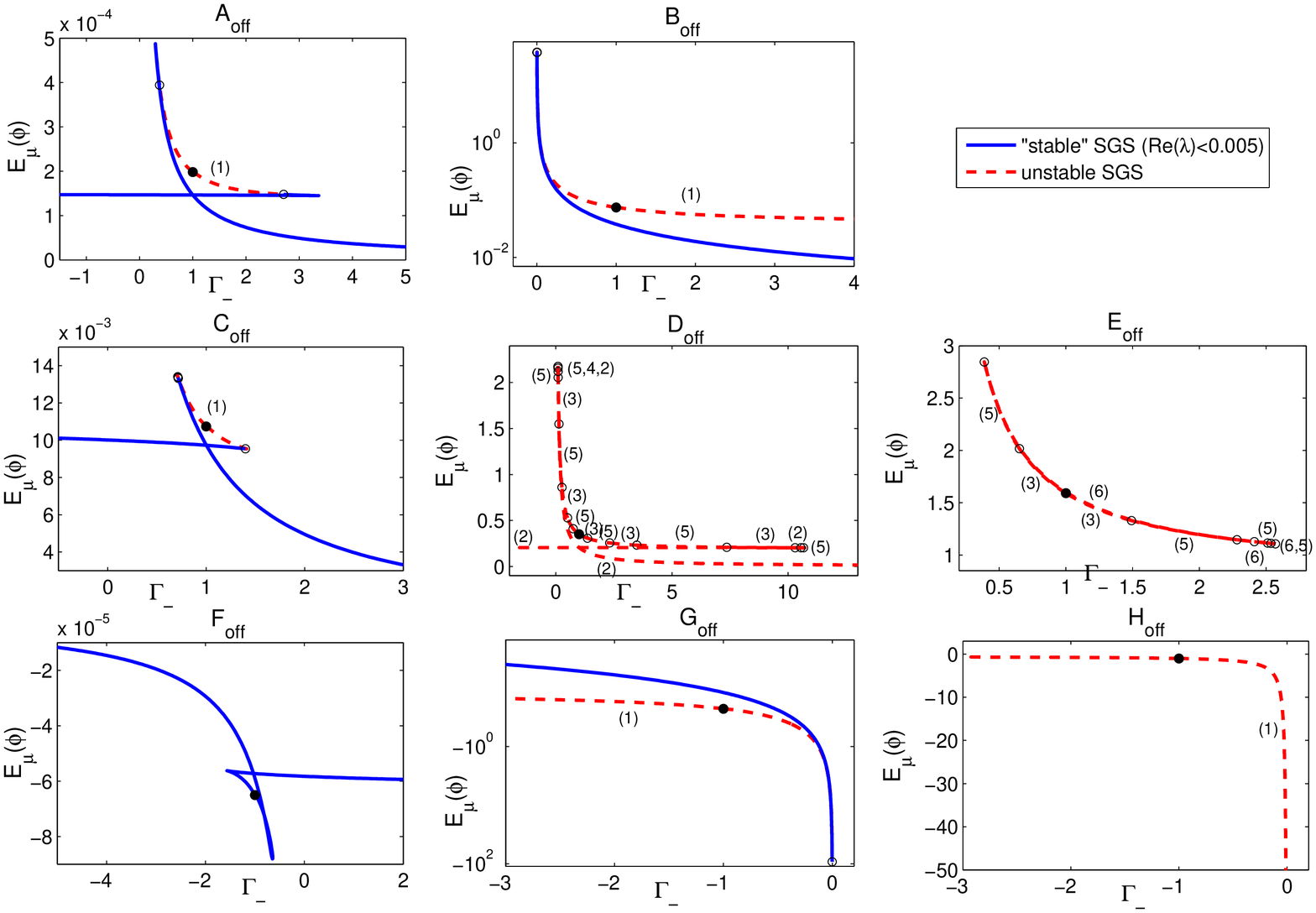,scale=.5}
\end{center}
\caption{Evans function winding number computations for selected SGS families bifurcating from offsite GSs. The vertical contour $\gamma$ with $\text{Re}(\gamma)=0.005$ was used. Full (blue) lines mark `stable' SGSs, where $n(E(\gamma),0)=0$ and dashed (red) lines unstable SGSs, where $n(E(\gamma),0)\neq 0$. Changes in the values of $n$ are marked by black circles and the value is written next to each segment of the curve.}
\label{F:SGS_off_stab}
\end{figure}

\begin{figure}[ht!]
\begin{center}
\epsfig{figure = 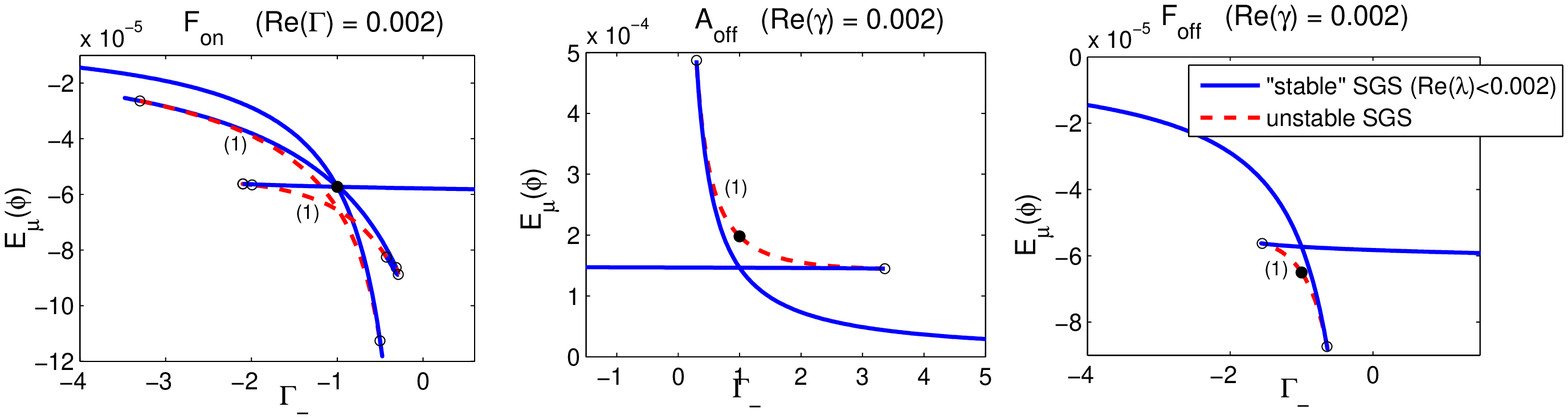,scale=.5}
\end{center}
\caption{Results of Evans function winding number computations for the SGS families $F_\text{on}$, $A_\text{off}$ and $F_\text{off}$ using the vertical contour $\gamma$ with $\text{Re}(\gamma)=0.002$. Apparently, a better match between the location of folds and stability changes is achieved than in the case $\text{Re}(\gamma)=0.005$, see Figs. \ref{F:SGS_on_stab} and \ref{F:SGS_off_stab}.}
\label{F:SGS_stab_near_im}
\end{figure}

For an SGS along the family $A_\text{off}$ Fig. \ref{F:Evans_fn_plot} shows the curve $E(\gamma)$ extended symmetrically about the real axis using Lemma \ref{L:E_sym}. The bold segments along $E(\gamma)$ in part (d) correspond to segments of the contour $\gamma$ lying close to $\sigma_\text{ess}(\L)$ as marked in (c). Clearly, in the vicinity of  $\sigma_\text{ess}(\L)$ the Evans function is more oscillatory than along the remaining parts of the contour.

For the selected SGS families from Fig. \ref{F:SGS_on} bifurcating from onsite GSs the results of winding number computations are in Fig. \ref{F:SGS_on_stab}. Similarly, Fig. \ref{F:SGS_off_stab} corresponds to the SGS families from Fig. \ref{F:SGS_off} bifurcating from offsite GSs. Full blue lines mark stable parts of the SGS families, where the winding number is zero, and the dashed red lines mark unstable parts, where the winding number is larger than $0$. The winding number values are written next to the curve and correspond to the adjacent segment demarcated by the black circles.

Except for the cases $A_\text{on}, F_\text{on}, A_\text{off}$ and $F_\text{off}$ each fold apparently marks a change in the value of the winding number. 
The reason why this does not happen for $A_\text{on}, F_\text{on}, A_\text{off}$ and $F_\text{off}$ is that in these cases all zeros of $E$ have small real parts and when the zeros lie between the imaginary axis and $\gamma$, the resulting winding number is zero. Fig. \ref{F:SGS_stab_near_im} shows three of these cases for the contour $\gamma$ with $\text{Re}(\gamma)=0.002$, i.e. closer to the imaginary axis. The folds now more closely match winding number changes, which in these cases all correspond to stability changes. We show, however, in Sec. \ref{S:eval_loc} computations suggesting that at least in the case of real eigenvalues bifurcations of eigenvalue from the imaginary axis happen near folds but not \textit{exactly} at folds.
%
%
In agreement with \cite{PSK04} we get that all off-site GSs (black dots in Fig. \ref{F:SGS_off_stab}) are unstable. For onsite GSs (black dots in \ref{F:SGS_on_stab}) we get `stability' ($\text{Re}(\gamma)< 5*10^{-3}$) except for $D_\text{on}$. Previously, see e.g \cite{LOSK03,AS05}, onsite GSs, including discrete GSs, have been demonstrated to be stable via direct numerical simulations of their evolution in time. As, however, shown in \cite{PSK04} onsite GSs can, in fact, be unstable when an eigenvalue bifurcates from an edge of $\sigma_\text{ess}(\L)$ away from the imaginary axis. This is caused by a resonance between the GS and an `internal mode'. From our selected cases $D_\text{on}$ is the only unstable one. Note that although the GS $C_\text{on}$ is plotted as stable in Fig. \ref{F:SGS_on_stab}, it is shown in Fig. 6 of \cite{PSK04} that it is unstable due to an eigenvalue $\lambda$ with $\text{Re}(\lambda)\approx 3.5*10^{-4}$. This eigenvalue lies, however, well to the left of our contour. We reproduce eigenvalues near this point in the next section.

\subsubsection{Determining Eigenvalue Locations}\label{S:eval_loc}

\begin{figure}[ht!]
\begin{center}
\epsfig{figure = 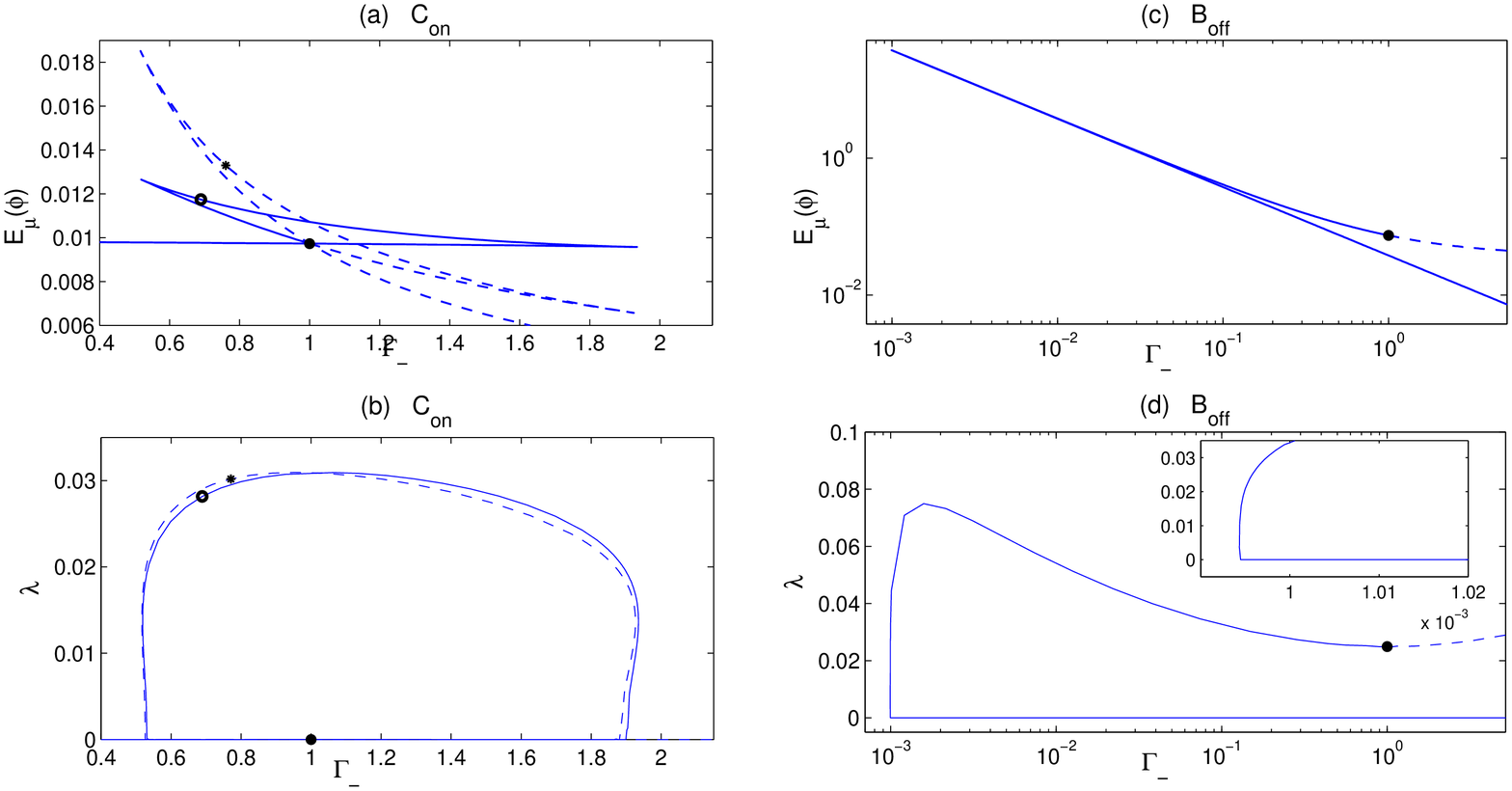,scale=.5}
\end{center}
\caption{The real eigenvalue for the families $C_{\text{on}}$ and $B_{\text{off}}$. Subplots (a) and (b) are for the family $C_{\text{on}}$ and (c) and (d) for $B_{\text{off}}$. (a,c): SGS continuation curves; (b,d): the real eigenvalue along the family. Markers along the curves are for a better orientation.}
\label{F:real_zeros}
\end{figure}

\begin{figure}[ht!]
\begin{center}
\epsfig{figure = 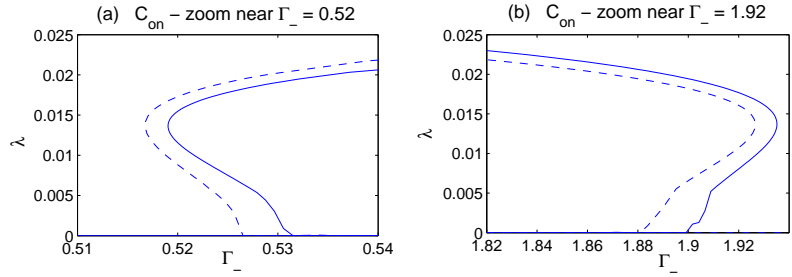,scale=.5}
\end{center}
\caption{A zoomed-in view of the bifurcations of real eigenvalues from zero for the family $C_{\text{on}}$.}
\label{F:real_zeros_zoom}
\end{figure}

Note that in general the method of the Evans function combined with the argument principle and a fixed contour cannot provide information on the strength of an instability since the precise location of eigenvalues is not known. When, however, the resulting winding number $n$ is odd, the symmetry in Lemma~\ref{L:E_sym} implies that at least one of the eigenvalues is real. Since in many of our examples $n=1$, the corresponding instability is exponential. In these cases the eigenvalue can be located by evaluating $E(\lambda)$ throughout $\lambda \in (0,\|\Gamma\|_\infty \|\phi_s\|_\infty^2)$ and interpolating the graph to obtain roots of $E$. The graphs of the root $\lambda$ along the SGS family $C_\text{on}$ and along the family $B_\text{off}$  are plotted in Fig. \ref{F:real_zeros} and \ref{F:real_zeros_zoom}. We see that in the case $C_\text{on}$ the eigenvalue does not bifurcate from $0$ precisely at the fold but shortly before the fold is reached, while for $B_\text{off}$ the bifurcation appears directly at the fold. 

\begin{figure}[ht!]
\begin{center}
\epsfig{figure = 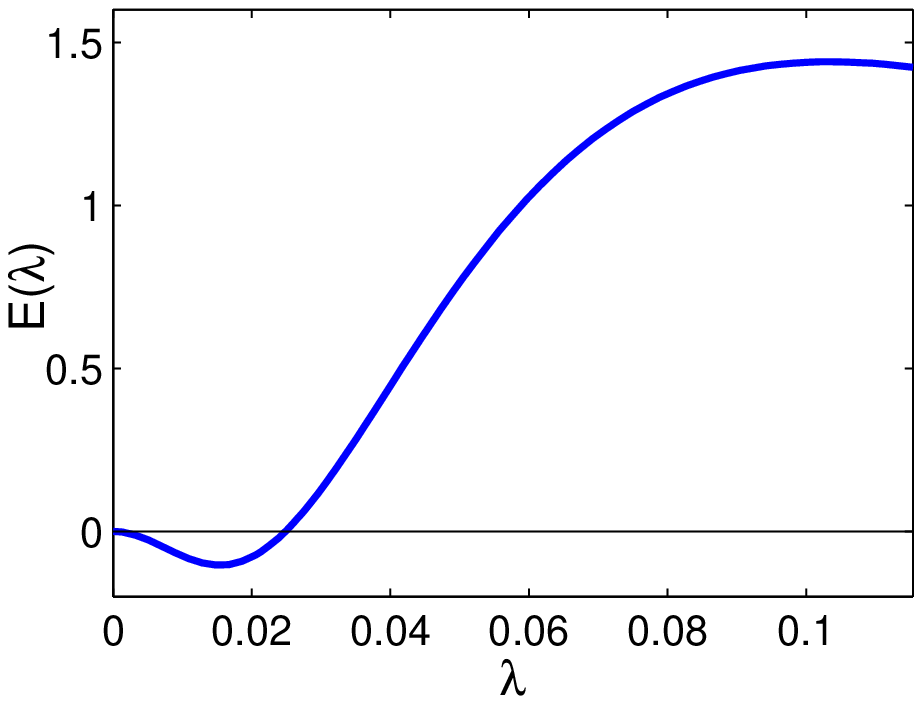,scale=.5}
\end{center}
\caption{The Evans function along $(0,\|\Gamma\|_\infty \|\phi_s\|_\infty^2)$ for the SGS at the circle marker in the family $C_\text{on}$ in Fig. \ref{F:real_zeros} (a).}
\label{F:E_along_R}
\end{figure}

For illustration in Fig. \ref{F:E_along_R} we show the plot of $E(\lambda)$ along $\lambda\in (0,\|\Gamma\|_\infty \|\phi_s\|_\infty^2)$ for the SGS at the circle marker in the family $C_\text{on}$ in Fig. \ref{F:real_zeros}.

In order to locate complex eigenvalues via the Evans function, there are, in principle, two alternatives: either a search algorithm employing the argument principle and contours enclosing each search region or a direct root finding approach based on a polynomial interpolation. We use the latter approach and select M\"{u}ller's method for finding complex zeros of complex functions \cite{Mueller56}. M\"{u}ller's method is a generalization of secant method in that instead of a linear interpolation it uses a quadratic one and thus requires three points for an initial guess. We use the implementation of D.H. Cortes \cite{Cortes08} and terminate M\"{u}ller's iteration if $|E(\lambda^{(k)})|< 10^{-12}$ and $|\lambda^{(k-1)}-\lambda^{(k)}|< 10^{-6}$ for the $k$-th iterate $\lambda^{(k)}$. Selected results have also been checked with Matlab's standard routine \verb fminsearch, which is based on a simplex search method.

\begin{figure}[ht!]
\begin{center}
\epsfig{figure = 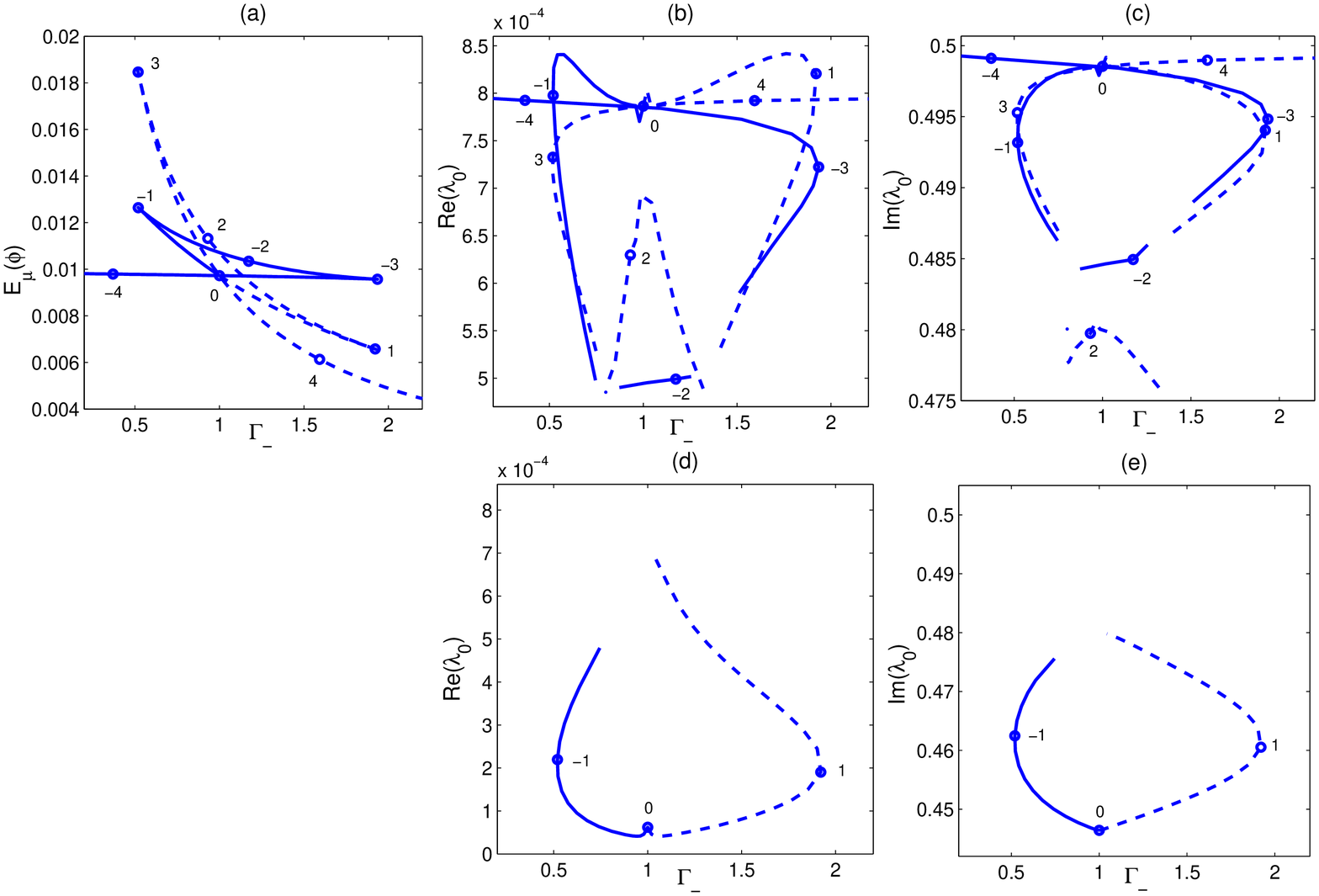,scale=.5}
\end{center}
\caption{Eigenvalue computation using M\"{u}ller's method. (a) SGS family $C_\text{on}$ with several points labeled; (b) The real part of the eigenvalues $\lambda_0$ obtained by tracking the eigenvalues of the gap soliton (labeled '0'). The gap soliton eigenvalue was computed with the starting guess $7.8*10^{-4}+0.498\ri$. (c) The corresponding imaginary part. (d) the same as (b) except the starting guess for the gap soliton is $10^{-4}+0.446*\ri$ (at the curve ends in (d) the eigenvalue switches to that plotted in (b)); (e) the corresponding imaginary part.}
\label{F:Con_mueller}
\end{figure}

We restrict our attention to eigenvalues in or near the slit $(0,0.005)+\ri \R$ omitted by the winding number computations in Sec. \ref{S:wind_nr_results} and select two examples, the families $C_\text{on}$ and $D_\text{off}$, for which we perform tracking of eigenvalues along the SGS families. 
For the $C_\text{on}$ gap soliton we select the starting guess triplet near $3.5*10^{-4}+0.49*\ri$, which was reported in Fig. 6 \cite{PSK04} to be an eigenvalue. After convergence we track the eigenvalue throughout the whole SGS family. The results are in Fig. \ref{F:Con_mueller}. Varying the initial guess we have found two eigenvalues for the gap soliton: $\lambda_0 \approx 7.8*10^{-4}+0.498\ri$ and $\lambda_0 \approx 0.5*10^{-4}+0.4447*\ri$, satisfying our tolerances. The former eigenvalue is tracked in Fig. \ref{F:Con_mueller} (b) and (c) and the latter one in (d) and (e). 

\begin{figure}[ht!]
\begin{center}
\epsfig{figure = 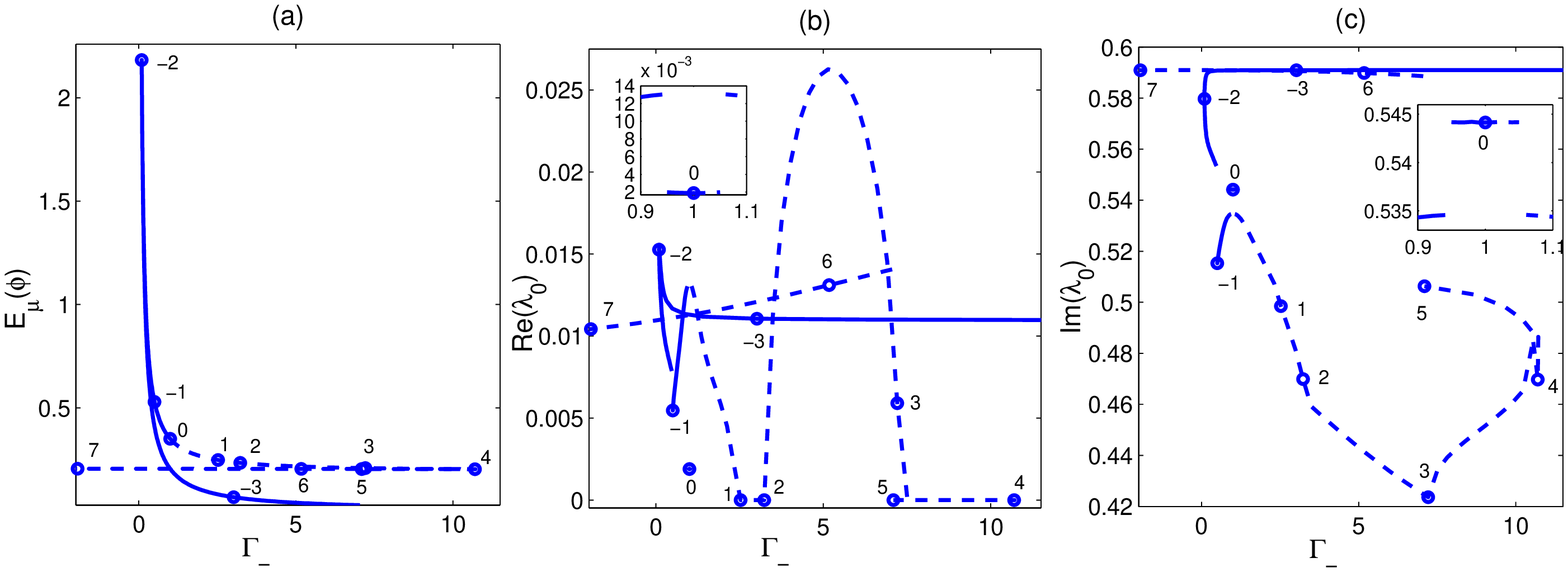,scale=.5}
\end{center}
\caption{Eigenvalue computation using M\"{u}ller's method. (a) SGS family $D_\text{off}$ with several points labeled; (b) The real part of the eigenvalues $\lambda_0$ obtained by tracking the eigenvalues of the gap soliton (labeled '0'). The gap soliton eigenvalue was computed with the starting guess $0.002+0.544*\ri$. (c) The corresponding imaginary part.}
\label{F:Doff_mueller}
\end{figure}

For the family $D_\text{off}$ and the starting guess triplet for the gap soliton near $0.002+0.544*\ri$ the computed eigenvalues are plotted in Fig. \ref{F:Doff_mueller}.

In both Fig. \ref{F:Con_mueller} and \ref{F:Doff_mueller} the eigenvalue curves have a number of discontinuities. At these points M\"{u}ller's iteration switches to another eigenvalue. The computations show that the slit $(0,0.005)+\ri \R$ missed by the winding number computations in many cases, indeed, contains additional eigenvalues. Numerically locating \textit{all} these eigenvalues is possibly unfeasible with existing methods.

\subsubsection{Direct Numerical PDE Evolution}
For several selected SGSs we check the stability results by direct numerical simulations of the $t$-dependent NLS \eqref{E:PNLS} using the corresponding SGS profile $\phi$ plus a small perturbation as the initial condition. In detail, we take
\[\psi(x,0)=\phi(x)+0.01 \max_{x}(\phi(x)) \sin^2\left({\pi x\over 2}\right)\sech\left({x\over 10}\right),\] 
where the second term represents an oscillatory perturbation with an expected large spectral support in $\sigma(\L)$ so that many modes of the linearization problem \eqref{E:evp} are excited. This perturbation is used merely to accelerate possible instabilities since the numerical approximation alone introduces error which has a large spectral support. The numerical integration of  \eqref{E:PNLS} was performed via a 4th order (in $dt$) split-step method \cite{Yoshida1990} where \eqref{E:PNLS} was rewritten as $\pa_t \psi = A\psi + B(x,\psi)\psi$ with $A = \ri \pa_x^2$ and $B(x,\psi) =-\ri V(x) +\ri \Gamma(x)|\psi(x)|^2$, so that splitting into the constant coefficient linear part $\pa_t\psi = A\psi$ and the nonlinear part $\pa_t \psi = B(x,\psi)\psi$ was applied. The former part was solved in Fourier $k-$space and the latter one in physical $x-$space exactly. In all simulations the computational box $x\in [-100, 100-dx]$ was used with $3000$ grid points so that $dx\approx 0.067$. The time step was $dt=0.01$. The two above mentioned conserved quantities of equation 
\eqref{E:PNLS}, namely the total power $\|\psi(\cdot, t)\|^2_{L^2(\R)}$ and the total energy $E_\mu(\psi)$, are preserved with our numerical method with a typical relative error around $10^{-5}$ and $10^{-4}$ respectively.
\begin{figure}[ht!]
\begin{center}
\epsfig{figure = 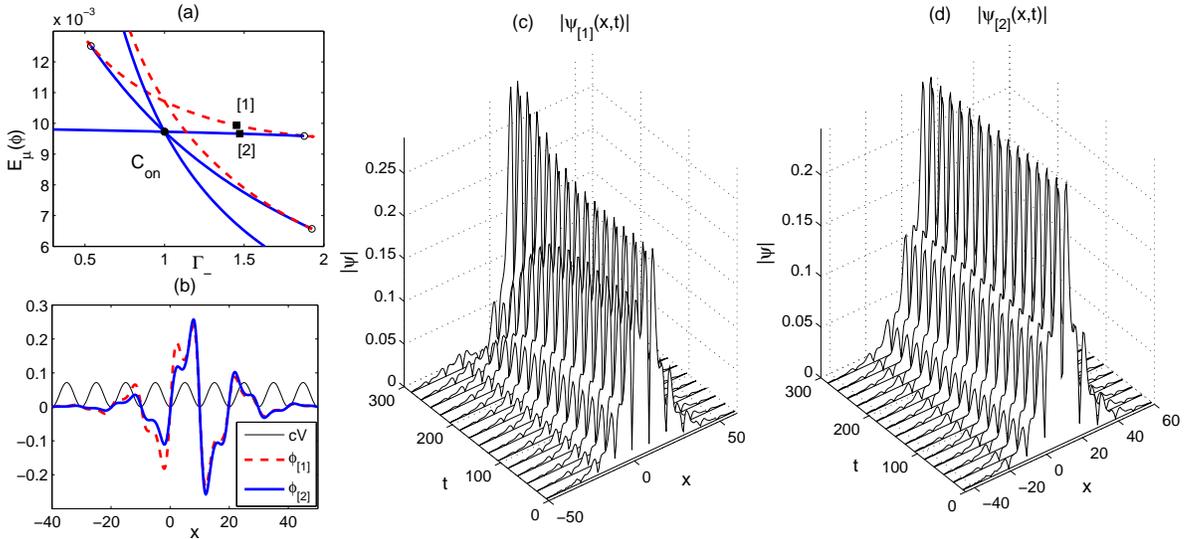,scale=.5}
\end{center}
\caption{Time evolution of 2 SGSs from the family $C_\text{on}$ \ ($\Gamma_+ = 1$). (a) the SGS family with stability marked as in Fig. \ref{F:SGS_on_stab}, and 2 selected labeled points: [1] with $\Gamma_- \approx 1.451$ and [2] with $\Gamma_- \approx 1.472$; (b) profiles of the 2 SGSs [1] and [2]; (c) and (d) evolution of the modulus $|\psi|$ for the 2 SGSs. Clearly, $\psi_{[1]}$ evolves unstably and $\psi_{[2]}$ stably in agreement with the stability diagram (a).}
\label{F:evol_C_on}
\end{figure}
\begin{figure}[ht!]
\begin{center}
\epsfig{figure = 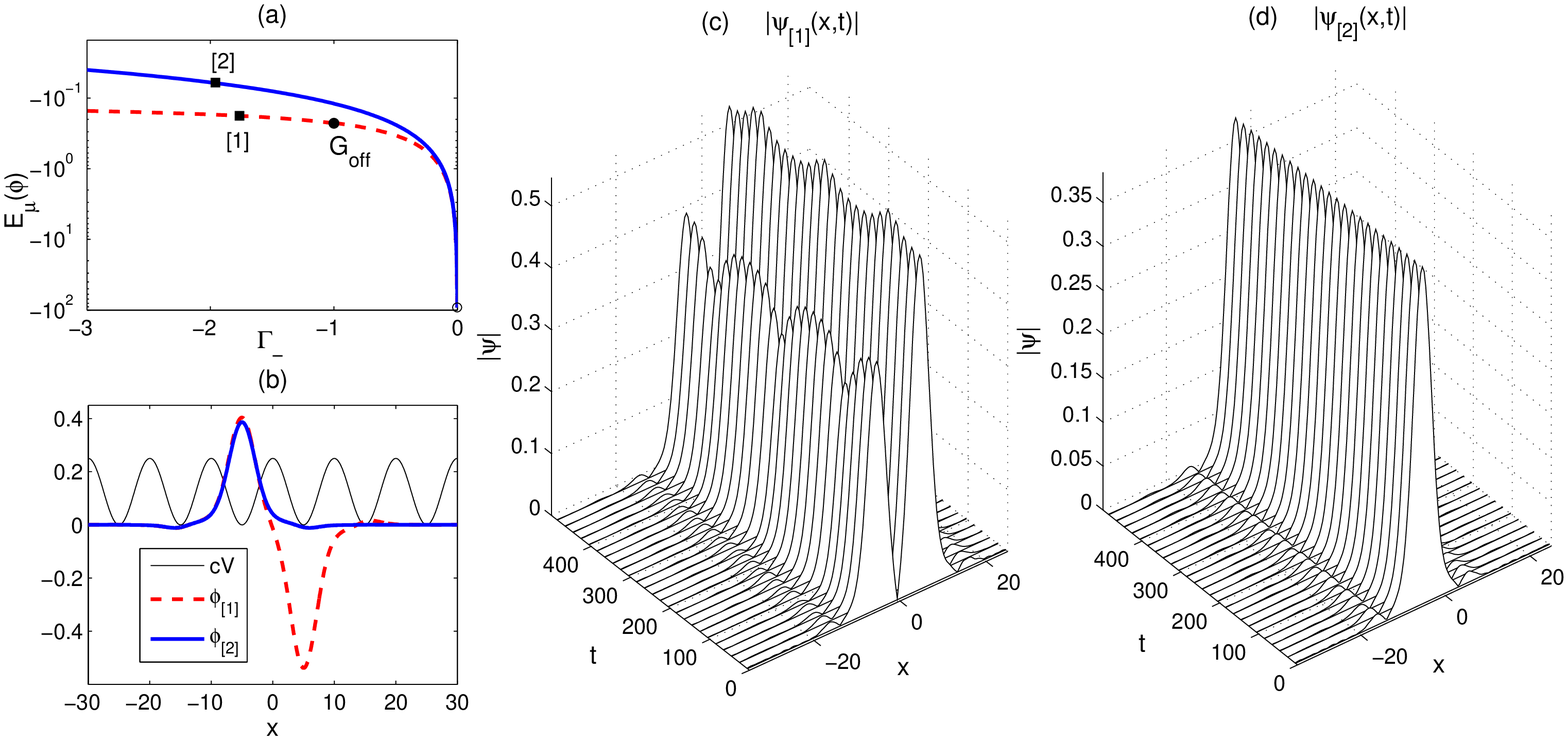,scale=.5}
\end{center}
\caption{Time evolution of 2 SGSs from the family $G_\text{off}$\ ($\Gamma_+ = -1$). (a) the SGS family with stability marked as in Fig. \ref{F:SGS_off_stab}, and 2 selected labeled points: [1] with $\Gamma_-\approx -1.764$ and [2] with $\Gamma_-\approx -1.959$; (b) profiles of the 2 SGSs [1] and [2]; (c) and (d) evolution of the modulus $|\psi|$ for the 2 SGSs. Clearly, $\psi_{[1]}$ evolves unstably and $\psi_{[2]}$ stably in agreement with the stability diagram (a).}
\label{F:evol_G_off}
\end{figure}
\begin{figure}[ht!]
\begin{center}
\epsfig{figure = 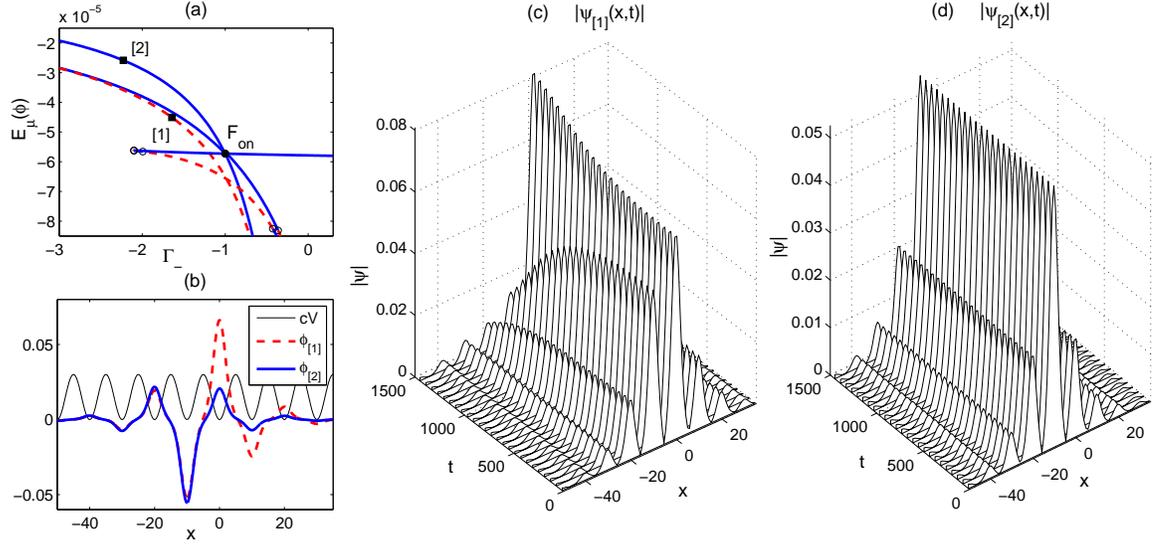,scale=.5}
\end{center}
\caption{Time evolution of 2 SGSs from the family $F_\text{on}$\ ($\Gamma_+ = -1$). (a) the SGS family with stability marked as in Fig. \ref{F:SGS_on_stab}, and 2 selected labeled points: [1] with $\Gamma_- \approx -1.643$ and [2] with $\Gamma_- \approx -2.229$; (b) profiles of the 2 SGSs [1] and [2]; (c) and (d) evolution of the modulus $|\psi|$ for the 2 SGSs. Clearly, $\psi_{[1]}$ evolves unstably and $\psi_{[2]}$ stably in agreement with the stability diagram (a).}
\label{F:evol_F_on}
\end{figure}

\begin{figure}[ht!]
\begin{center}
\epsfig{figure = 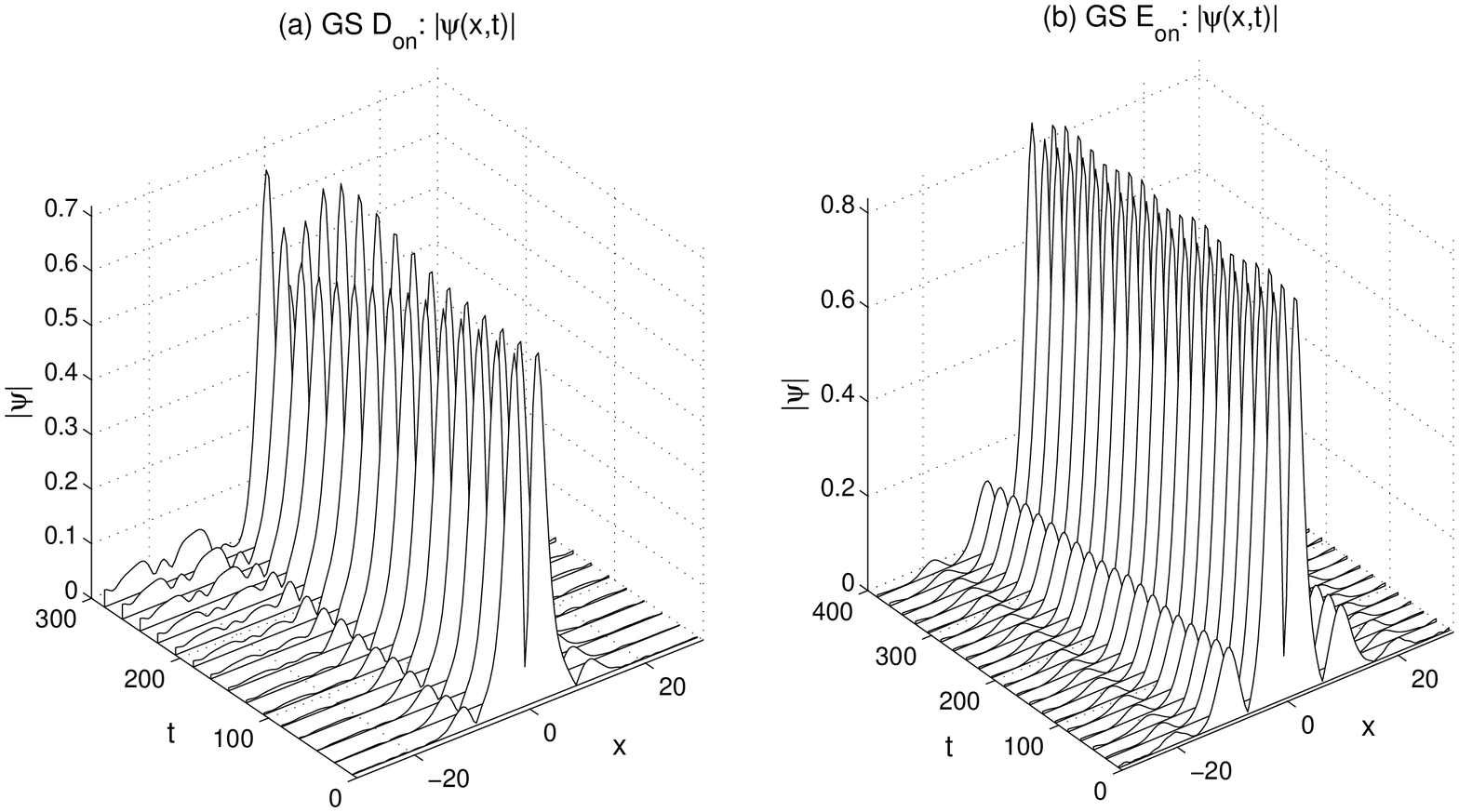,scale=.5}
\end{center}
\caption{Time evolution of the GSs $D_\text{on}$ and $E_\text{on}$ \ ($\Gamma_+ = \Gamma_- = 1$). In Fig. \ref{F:SGS_on_stab} Evans function computations show these to be unstable and stable respectively. The $t-$evolution agrees with these stability results.}
\label{F:evol_GS_Don_Eon}
\end{figure}

Figs. \ref{F:evol_C_on},\ref{F:evol_G_off} and \ref{F:evol_F_on} show the evolution of one stable and one unstable SGS along the families $C_\text{on}, G_\text{off}$ and $F_\text{on}$, respectively. 
And Fig. \ref{F:evol_GS_Don_Eon} shows the evolution of the onsite GSs ($\Gamma_-=\Gamma_+=1$) $D_\text{on}$ and $E_\text{on}$. In all cases the (in)stability of the dynamics is in agreement with the results in Figs. \ref{F:SGS_on_stab}, \ref{F:SGS_off_stab}, and \ref{F:SGS_stab_near_im}. In the case $F_\text{on}$ the time interval had to be chosen larger than in the other cases due to the weaker instability: for all the SGS in the $F_\text{on}$ family $\text{Re}(\lambda)<0.005$ while in the other families the unstable solutions have $\text{Re}(\lambda)>0.005$ as follows from Figs. \ref{F:SGS_on_stab}, \ref{F:SGS_off_stab}, and \ref{F:SGS_stab_near_im}.

\begin{figure}[h!]
\begin{center}
\epsfig{figure = 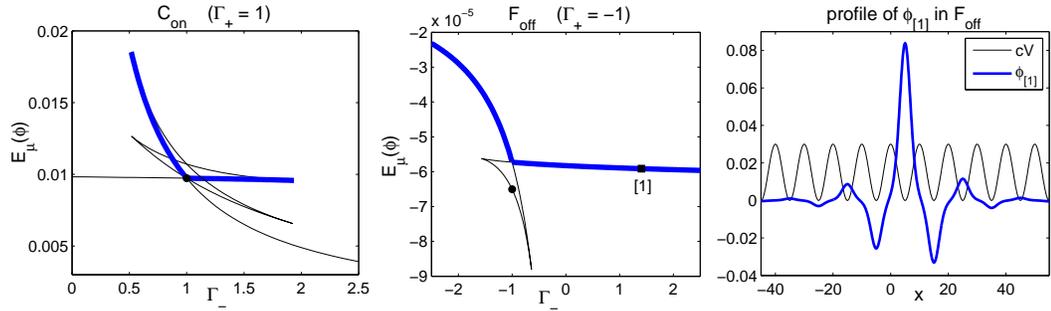,scale=.5}
\end{center}
\caption{Left and middle: "stable" segments ($\text{Re}(\lambda)<0.005$) of the SGS families $C_\text{on}$ (where $\Gamma_+=1$) and $F_\text{off}$ (where $\Gamma_+=-1$), which consist of solutions centered in the less focusing part of the domain are marked in bold. On the right: profile of the "stable" SGS $(1)$ at $\Gamma_- \approx 1.41$ marked along the $F_\text{off}$ family. This SGS is at a truly focusing/defocusing interface and concentrated mostly in the defocusing half.}
\label{F:stab_less_foc}
\end{figure}

Based on a physical intuition from optics one might expect that SGSs with their center in the more defocusing half of the medium (i.e. the half with the smaller value of $\Gamma$) should be unstable due to the tendency of light to converge to the more focusing regions. Mathematically this expectation also makes sense since the total energy \eqref{E:energy} is decreased via a shift toward the more focusing region. This is, however, only a heuristic view due to the lack of shift invariance in the system. In reality nothing, in general, prevents the possibility of existence of a local extremum of the energy with the extremizer localized in the less focusing region and being stable. As we can see, for instance, in the SGS families $C_\text{on}$ and $F_\text{off}$ by inspecting Figs. \ref{F:SGS_vert_line} and \ref{F:SGS_on_stab}, there are parts of the family where this intuition indeed fails. The segments of the families marked in Fig. \ref{F:stab_less_foc} correspond to stable SGSs centered in the less focusing half. Note that for the segment of the $F_\text{off}$ family exiting the plot window on the right the part that satisfies $\Gamma_->0$ is a family of "stable" solutions ($\text{Re}(\lambda)<0.005$) at a focusing/defocusing interface and centered in the defocusing medium!
Similar segments of families of SGSs centered in the less focusing half can be found, for instance, in the cases $A_\text{on}, F_\text{on}, A_\text{off}$, and $C_\text{off}$.

\section{Conclusions}

Surface gap solitons (SGSs) of the cubic 1D periodic Schr\"{o}dinger (PNLS) / Gross Pitaevskii (GP) equation with a nonlinearity interface have been studied. As the computations show, bright SGSs exist even in the case of a focusing/defocusing interface. In \cite{DP08} SGSs were constructed via a numerical parameter continuation from gap solitons (GSs) and the families were followed up to the first fold. Here we have extended these families using the numerical arclength continuation. As shown, the SGS families often contain GSs centered at different extrema of the linear periodic potential or several non-identical GSs at the same extremum. For families of spatially broad SGSs the number of folds appears to equal the number of shifts of the SGS center of mass from one extremum of the periodic potential to the next. Asymptotic methods reveal that for SGSs bifurcating from GSs centered arbitrarily far from from the interface the allowed size of the jump in the nonlinearity coefficient decays as the propagation constant/frequency approaches the corresponding bifurcation edge of the spectral gap. Numerical arclength continuation results confirm this statement.

To inspect linear stability of SGSs we have developed a numerical Evans function method. Eigenvalues with the real part to the right of a small threshold have been detected via the complex argument principle applied to the Evans function with a vertical contour near the imaginary axis. This is the first time, to our knowledge, that the Evans function method combined with the argument principle has been applied to a problem with periodic coefficients. Standard difficulties in evaluating the Evans function in a numerically stable way have been reported and explained in detail in the context of the problem at hand. These include firstly stiffness of the ODE linearization problem on each of the unstable and stable manifolds, which need to be approximated for the Evans function evaluation. This is overcome via a reformulation in exterior algebra. Secondly, a numerical approximation of solutions with a fast exponential growth is necessary, which is performed under a change of variables. Next, accurate winding number computations require an adaptive discretization of the contour. Finally, the accuracy of the Evans function 
is shown to depend on the level of preservation of the weak Grassmanian invariant in the evolution of the two manifolds and on the discretization error in the solution of the SGS profile.

The results reveal the existence of both unstable and stable SGSs, where stability does not exclude the presence of eigenvalues with a positive real part below the threshold given by the distance of our contour to the imaginary axis. The results on stability of GSs are in agreement with the previous results in \cite{PSK04}. Interestingly, stability is possible also for SGSs centered in the less focusing medium of the two and even when the SGS is concentrated in the defocusing half with a focusing-defocusing interface. Direct numerical simulations of the PNLS/GP equation were performed for several SGS examples and confirm the (in)stability statements based on Evans function computations.

An apparent shortcoming of the argument principle approach is the lack of information on eigenvalues between the imaginary axis and the vertical contour. Note that the imaginary axis cannot be chosen as the contour even if the Evans function $E$ is analytically extended through the essential spectrum, see \cite{KS98,KS02,KS04} for an analytic extension in constant coefficient systems, due to the occurrence of zeros of $E$ on the imaginary axis. 
In order to gain insight on the region  between the imaginary axis and the vertical contour, we have tracked several complex eigenvalues there via the use of M\"{u}ller's method. The authors plan to address numerically eigenvalue bifurcations from the essential spectrum by using an analytic extension of the Evans function in a future paper.

All the techniques presented in this paper can be easily adapted also for problems with linear interfaces, e.g. when the linear potential has the form $V(x) = V_{+}(x) \; \chi_{[0,\infty)}(x) + V_{-}(x) \; \chi_{(-\infty,0)}(x)$ with periodic $V_\pm$. In fact, only periodicity of $V$ for large enough $|x|$ is a requirement for the techniques. The stability method can thus be directly applied, for instance, to the linear interface SGSs in \cite{KVT06,MHChSMS06}. In addition the form of the nonlinearity coefficient can be arbitrary. A practical advantage of the numerical Evans function method is the possibility of a straightforward parallelization of the computations by dividing the contour into segments.  

Computation of SGSs via the arclength continuation generalizes directly to more spatial dimensions. The application of the Evans function is, however, generally unfeasible as the stable and unstable manifolds are typically infinite dimensional in more than one spatial dimension.  

\section{Acknowledgments}
The authors would like to thank Markus Richter, Karlsruhe Institute of Technology, for providing his FEM package with which SGS profiles were computed.
Further, we want to thank Wolfgang Reichel, Karlsruhe Institute of Technology,  Dmitry Pelinovsky, McMaster University, and Bj\"{o}rn Sandstede, Brown University for numerous helpful discussions and suggestions.

\appendix
\section{Winding Number Code}\label{A:wind_code}
For readers' convenience and to motivate the use of winding number based computations of point spectra by others, we provide here our Matlab code for the computation of the winding number of a closed curve in the complex plane with respect to the point $0$. A remark on the choice of a starting point is included at the beginning of the code.
\begin{lstlisting}
function wind = wind_nr(E)

%%%%%%%%%%%%%%%%%%%%%%%%%%%%%%%%%%%%%%%%%%%%%%%%%%%%%%%%%%%%%%%%%%%%%%%%%%%%%%%
% winding number computation
% input: complex array E of the size [N x 1] representing a closed curve
%        in the complex plane ( E(1) = E(N) )
% output: winding number wind of the curve with respect to the point 0

% beware: choice of ind_st
%%%%%%%%%%%%%%%%%%%%%%%%%%%%%%%%%%%%%%%%%%%%%%%%%%%%%%%%%%%%%%%%%%%%%%%%%%%%%%%

plots_on=1;
if(plots_on)
    figure(1); plot(E)
    xlabel('Re(E)'); ylabel('Im(E)')
end

%select a starting point for the winding number computation
%(should not be at a point where the curve E is tangent to a ray from the origin
% or where the phase is \pm \pi)
ind_st=3;
%for the Evans function example with an unbounded lambda-contour where the
%first m and the last m entries of E are a line connecting the last computed value
%of E along the lambda-contour and the asymptotic value 1 of E for
%|lambda|-> oo a suitable choice of ind_st is 1 < ind_st < m because the complex 
%phase of E changes monotonically and the phase value is not +/- pi for 1<ind_st<m


%initiate the value of the phase and the winding number
phase_0 = atan2(imag(E(ind_st)),real(E(ind_st)));
phase = atan2(imag(E(ind_st+1)),real(E(ind_st+1)));
if(ind_st==1)
    phase_old = atan2(imag(E(end-1)),real(E(end-1)));
else
    phase_old = atan2(imag(E(ind_st-1)),real(E(ind_st-1)));
end

if(phase>phase_0 && phase_0>phase_old && abs(phase-phase_old)<pi )
    wind = 1;
elseif(phase<phase_0 && phase_0<phase_old && abs(phase-phase_old)<pi )
    wind = -1;
end

%run along the E-curve tracking the phase and increasing or decreasing
%the winding number counter
check_monot = 0;
phase_old = phase_0;
for k = [ind_st+1:length(E) 2:ind_st]
    phase=atan2(imag(E(k)),real(E(k)));
    
    if(check_monot==1) %check if E is tangent to the ray with the angle phase_0
        if(phase > phase_old && phase_old_old < phase_old && abs(phase-phase_old)<pi)
            wind = wind + 1;
        elseif(phase < phase_old && phase_old_old > phase_old && abs(phase-phase_old)<pi)
            wind = wind - 1;
        end
        check_monot = 0;
    end
    
    %check if the value phase_0 has been passed
    %(i.e. if E has crossed the ray with the angle phase_0)
    if(phase>phase_0 && phase_old<phase_0 && abs(phase-phase_old)<pi)
        %counter clock-wise crossing
        wind = wind + 1;
    elseif(phase<phase_0 && phase_old>phase_0 && abs(phase-phase_old)<pi)
        %clock-wise crossing
        wind = wind - 1;
    elseif(phase==phase_0) %need to check if phase has an extremum here
        %(i.e. if E is tangent to the ray with the angle phase_0)
        phase_old_old = phase_old;
        check_monot = 1;
    end
    phase_old = phase;
end
end
\end{lstlisting}

\bibliographystyle{siam}
\bibliography{bibliography}

\end{document}